\documentclass[sigconf,edbt]{acmart-edbt2019}
\usepackage[T1]{fontenc}
\usepackage[utf8]{inputenc}
\usepackage[bottom]{footmisc}
\usepackage{booktabs} 
\usepackage{float}
\usepackage{subfig}
\usepackage{silence}
\usepackage{tabularx}
\usepackage{etoolbox}
\usepackage{MnSymbol}
\usepackage{hyperref}
\usepackage{amsmath}
\usepackage{listings}
\usepackage{fancybox}
\usepackage{wrapfig}
\usepackage{graphicx}
\usepackage{url}
\usepackage{paralist}
\usepackage{multirow}
\usepackage{balance}  

\newcommand{\qq}[1]{``#1''}

\usepackage[normalem]{ulem}
\usepackage{lipsum}

\usepackage{enumitem}
\setlist[enumerate]{itemsep=0mm}

\newcommand{\mycodefont}{
\fontsize{8}{9}\selectfont\ttfamily
}

\newcommand{\code}[1]{\texttt{#1}}
\graphicspath{{figures/}}

\setcopyright{rightsretained}

\acmDOI{}

\acmISBN{XXX-X-XXXXX-XXX-X}

\acmConference[XYZ 2019]{XYZ}{March XX-YY, 2019}{XYZ,XYZ} 
\acmYear{2019}

\begin{document}


\title{No One is Perfect: Analysing the Performance of Question Answering Components over the DBpedia Knowledge Graph}

\author{Kuldeep Singh}
\affiliation{%
  \institution{Fraunhofer IAIS}
  \city{Sankt Augustin, Germany}
}
\email{kuldeep.singh@iais.fraunhofer.de}

\author{Ioanna Lytra}
\affiliation{%
  \institution{University of Bonn \& Fraunhofer IAIS}
  \city{Bonn, Germany}
}
\email{lytra@cs.uni-bonn.de}

\author{Arun Sethupat Radhakrishna}
\affiliation{%
  \institution{University of Minnesota}
  \city{Minnesota, USA}
}
\email{sethu021@umn.edu}

\author{Saeedeh Shekarpour}
\affiliation{%
  \institution{University of Dayton}
  \city{Dayton, USA}
}
\email{sshekarpour1@udayton.edu}

\author{Maria-Esther Vidal}
\affiliation{%
  \institution{TIB \& L3S Research Centre}
  \city{Hannover, Germany}
}
\email{lytra@cs.uni-bonn.de}

\author{Jens Lehmann}
\affiliation{%
  \institution{University of Bonn \& Fraunhofer IAIS}
  \city{Bonn, Germany}
}
\email{jens.lehmann@cs.uni-bonn.de}

\renewcommand{\shortauthors}{}

\begin{abstract}
Question answering (QA) over knowledge graphs has gained significant momentum over the past five years due to the increasing availability of large knowledge graphs and the rising importance of question answering for user interaction. 
DBpedia has been the most prominently used knowledge graph in this setting and most approaches currently use a pipeline of processing steps connecting a sequence of components.
In this article, we analyse and micro evaluate the behaviour of 29 available QA components for the DBpedia knowledge graph that were released by the research community since 2010. As a result, we provide a perspective on collective failure cases, suggest characteristics of QA components that prevent them from performing better and provide future challenges and research directions in the field. 
\end{abstract}

\maketitle

\section{Introduction} \label{sec:introduction}
\begin{figure*}[t]
	\centering
	\includegraphics[width=.97\textwidth]{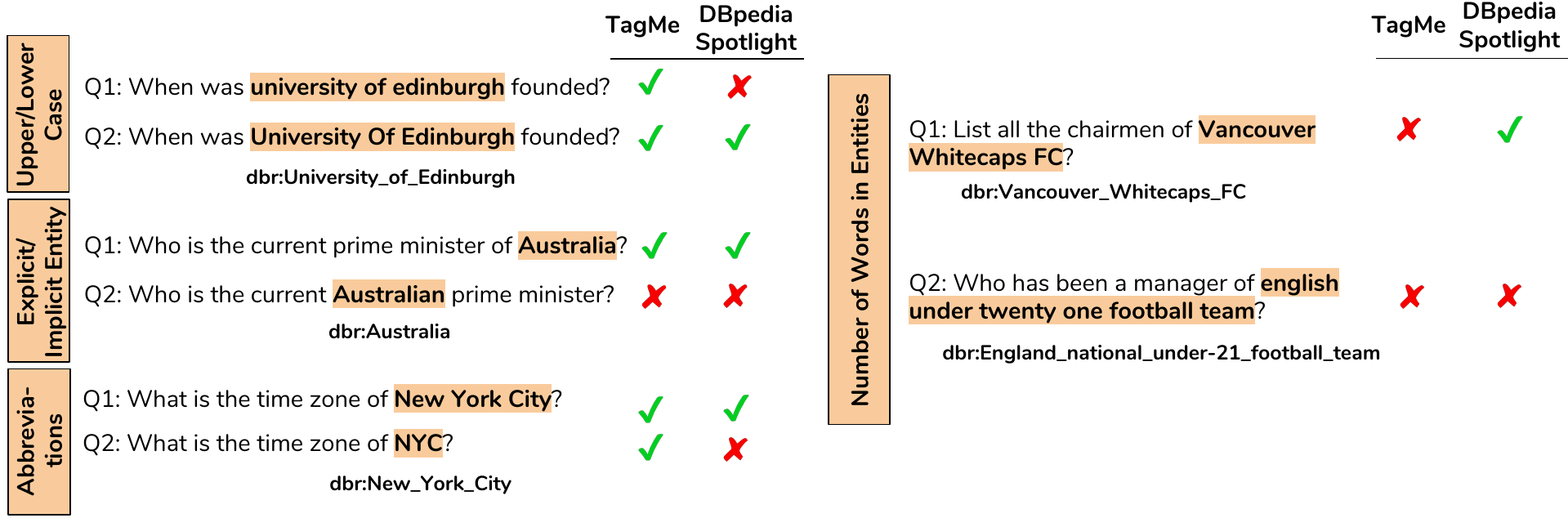}
	\caption{
	Performance of Top-2 NED Components on Specific Questions. TagMe and DBpedia Spotlight are the top-2 NED components in average on the LC-QuAD dataset (as shown in \autoref{tab:motivating}). However, their behaviour varies with regard to question features, for which entities in lowercase vs.\ uppercase, including abbreviations, having implicit vs.\ explicit mappings and with varying number of words have to be mapped to the DBpedia knowledge graph entities. 
	}
	\label{fig:motivating}
\end{figure*}
In the era of Big Knowledge, Question Answering (QA) systems allow for responding natural language or voice-based questions posed against various data sources, e.g.~knowledge graphs, relational databases or documents~\cite{DBLP:journals/pvldb/LiJ14,DBLP:journals/pvldb/CuiXWSHW17}. 
Particularly, with the advent of open knowledge graphs (e.g.~DBpedia~\cite{DBLP:conf/semweb/AuerBKLCI07}, Freebase~\cite{DBLP:conf/aaai/BollackerCT07} and Wikidata~\cite{DBLP:conf/www/Vrandecic12}), question answering over structured data gained momentum and researchers from different communities, e.g.~semantic web, information retrieval, databases and natural language processing have extensively studied this problem over the past decade~\cite{GERBILQA,DBLP:conf/edbt/MazzeoZ16,DBLP:conf/sigmod/ZouHWYHZ14}. Thus, since 2010, more than 62 QA systems have been published, and DBpedia is the underlying knowledge graph in 38 of them~\cite{GERBILQA}.
Those systems usually translate natural language questions to a formal representation of a query that extracts answers from the given knowledge graph.
The analysis of the architecture of these QA systems over DBpedia shows that the QA system architectures share similar tasks on the abstract level. These tasks include Named Entity Recognition and Disambiguation (NER and NED), Relation Linking (RL), Class Linking (CL), answer type identification, dependency parsing and Query Building (QB)~\cite{diefenbach2017core}. 

For instance, for the question \qq{What is the time zone of New York City?}, an ideal QA system over DBpedia generates a formal representation of this question, that is a formal query (here expressed as SPARQL\footnote{\url{https://www.w3.org/TR/rdf-sparql-query/}}), which retrieves answers from the DBpedia endpoint\footnote{\url{http://dbpedia.org/sparql}} (i.e.~\texttt{SELECT ?c \{dbr:New\_York\_City dbo:timeZone ?c.\}}).
During this process, a QA system performs successive QA tasks. In the first step (i.e.~NED), a QA system is expected to recognise and link the entity present in the question to its candidate(s) from DBpedia (i.e.~mapping \texttt{New York City} to \texttt{dbr:New\_York\_City}\footnote{Prefix \code{dbr} is bound to \code{http://dbpedia.org/resource/}}). The next step is RL, in which QA systems link the natural language predicate to the corresponding predicate in DBpedia (i.e. mapping \texttt{time zone} to \texttt{dbo:timeZone}\footnote{Prefix \code{dbo} is bound to \code{http://dbpedia.org/ontology/}}). Finally, the QB component obtains the linked resources and predicates from the previous steps to formulate a SPARQL query.

\noindent
{\bf Research Objective.} \begin{inparaenum}[\itshape i\upshape)]
Several independent QA components for various QA tasks (e.g. NED and RL) have been released by the research community. These components are earlier reused in QA frameworks such as openQA~\cite{openqa}, QALL-ME~\cite{qallme}, OKBQA~\cite{kim2016open} and Frankenstein~\cite{DBLP:conf/esws/SinghBRS18} to build QA systems in collaborative community efforts rather building a system from scratch. Recent empirical studies have revealed that albeit overall effective, the performance of monolithic QA systems and QA components depends heavily on the features of input questions~\cite{saleemquestion,DBLP:conf/www/SinghRBSLUVKP0V18}, and not even the combination of the best performing QA systems or individual QA components retrieves complete and correct answers~\cite{DBLP:journals/pvldb/CuiXWSHW17}. In order to advance the state of the art in building QA systems and explore future research directions, it is important to get insights into the strengths and weaknesses of the range of existing QA components that can be reused in building QA systems. \textcolor{black}{Therefore, the goal of this evaluation study is to analyse the performance of existing reusable QA components implementing the aforementioned QA tasks with respect to different question features, and identify, thus, common behaviours, collective failures and directions for future QA research.}
\end{inparaenum}
~\\
{\bf Approach.} \begin{inparaenum}[\itshape i\upshape)]
In this paper, we aim at putting main properties of existing QA approaches over DBpedia in perspective and provide strong evidence and conclusive insights about the parameters that affect the performance of state-of-the-art QA approaches. 
We have collected 29 QA components that implement various QA tasks, and evaluate each component over 3,000 questions from the LC-QuAD dataset~\cite{trivedi2017lc}. 19 out of 29 QA components are accompanied by peer reviewed publications, and other 10 components are openly available to be reused for building QA systems. These components are evaluated using the questions range from simple to complex and vary in the number of entities and relations as well as expressiveness. Overall, 59 variables or question features are observed during the evaluation of these questions in the 29 QA components. The outcome of this evaluation uncovers characteristics of the studied QA components that allow for the explanation of their diverse behaviour. First, there are certain questions that none of the components or their combinations are able to answer, e.g. all NED components fail to answer some questions with different capitalisation. Second, we also have observed that best performing QA components for a particular QA task are not the best for all the question features. For example, TagMe~\cite{DBLP:conf/cikm/FerraginaS10} is the overall best component evaluated by Singh et al.~\cite{DBLP:conf/www/SinghRBSLUVKP0V18} for the NED task, but it is not the best component for all the question features reported in this paper.
\end{inparaenum}
~\\
{\bf Contributions.} \begin{inparaenum}[\itshape i\upshape)]
Our work contributes with four folds: First, results and analysis from a micro evaluation of 29 QA components based on 59 question features using over 3,000 questions per component are presented. Second, we provide a perspective on collective failure cases of state-of-the-art QA components and pitfalls of these components. Third, we collect insights suggesting the characteristics of the QA components that prevent them from correctly and completely performing their corresponding QA task. Finally, challenges and research directions required to be addressed in order to advance the state of the art are discussed.
\end{inparaenum}
~\\
The remainder of the paper is structured as follows. We illustrate the diverse behaviour exhibited by existing QA components in \autoref{sec:motivation}. Then, \autoref{sec:relatedwork} discusses related work and analyses drawbacks of existing QA benchmarkings and evaluations. In \autoref{sec:setting}, the experimental configuration is detailed, and the results of the empirical evaluation are reported in \autoref{sec:results}. We analyse the collective failures of QA components in \autoref{sec:failure}. In \autoref{sec:discussion}, we discuss and provide insights about the observed results, and \autoref{sec:conclusions} concludes with an outlook on future work.

\section{Motivating Examples} \label{sec:motivation}
One of the major tasks in a QA system is Named Entity Disambiguation (NED). 
Although there is a high number of implementations for the NED task~\cite{DBLP:conf/cikm/FerraginaS10,DBLP:conf/emnlp/HoffartYBFPSTTW11,MendesJGB11}, none of the implementations consistently performs best on a variety of input questions.
Various aspects of the input questions, e.g. length or complexity, might influence the performance of NED components.
Our motivating example demonstrates the dependency of the performance of two well-known NED implementations for various expressions of natural language input questions (cf.\ \autoref{fig:motivating}). For example, with respect to the question \qq{When was Edinburgh University founded?}, an ideal NED component targeting DBpedia\footnote{\url{https://wiki.dbpedia.org/}}~\cite{DBLP:conf/semweb/AuerBKLCI07} as underlying knowledge graph will identify the segment \texttt{Edinburgh University} as a named entity and afterwards link it to suitable candidate(s) in the knowledge graph, e.g. \texttt{dbr:University\_of\_Edinburgh}. TagMe\footnote{\url{https://services.d4science.org/web/tagme/demo}}~\cite{DBLP:conf/cikm/FerraginaS10} and DBpedia Spotlight\footnote{\url{https://www.dbpedia-spotlight.org/demo/}} \cite{MendesJGB11} are NED components evaluated in~\cite{DBLP:conf/www/SinghRBSLUVKP0V18} as the top 2 (out of 18) components, assessed using the LC-QuAD~\cite{trivedi2017lc} dataset. The precision, recall and F-score for these components are reported in \autoref{tab:motivating}. 
In the following paragraphs, we show the dependency of their performance on the input question.

\begin{table}[hb!]
	\centering
		\caption{Top-2 NED Components on DBpedia. The performance of TagMe and DBpedia Spotlight is reported based on the LC-QuAD dataset. They are the best performing components across 3,253 questions; however, the performance of these NED components varies depending on the question features (cf.\ \autoref{fig:motivating}).}
	\resizebox{0.96\columnwidth}{!}{%
        \begin{tabular}{ l l l l l }
    	    \toprule
            \textbf{QA Component} & \textbf{QA Task} & \textbf{Precision} & \textbf{Recall} & \textbf{F-score} \\
            \midrule
            {\it TagMe }
                & NED & 0.69 & 0.66 & 0.67 \\
            {\it DBpedia Spotlight}
                & NED & 0.58 & 0.59 & 0.59 \\
            \bottomrule
        \end{tabular}
        }
    \label{tab:motivating}
\end{table} 
~\\
{\bf Impact of Uppercase/Lowercase Entities.} \begin{inparaenum}[\itshape i\upshape)]
For the question \qq{When was Edinburgh University founded?}, TagMe and DBpedia Spotlight are able to successfully recognise and disambiguate the entity \texttt{dbr:University\_of\_Edinburgh}. If the given question is rephrased to \qq{When was university of edinburgh founded?}, in which the entity mention is written in \textit{lowercase letters}, TagMe continues to identify and disambiguate the correct entity whereas DBpedia Spotlight fails to do so. However, if the question is further rephrased to \qq{When was University Of Edinburgh founded?}, in which the entity segment appears in \textit{uppercase letters}, both NED components successfully disambiguate the entity.
\end{inparaenum}
~\\
{\bf Impact of Implicit/Explicit Entities.} \begin{inparaenum}[\itshape i\upshape)]
Let us consider another question, namely, \qq{Who is the current prime minister of Australia?}. In this question, the correct DBpedia mention for the named entity \texttt{Australia} is \texttt{dbr:Australia}. TagMe and DBpedia Spotlight can correctly disambiguate this entity. However, if the question is rephrased to \qq{Who is the current Australian prime minister?}, both components fail to link the entity \texttt{Australian} to \texttt{dbr:Australia}.
This is due to the fact that the entity is no longer an exact match of its DBpedia label \texttt{dbr:Australia} (i.e. vocabulary mismatch problem \cite{DBLP:conf/aaai/ShekarpourMAS17}), and therefore, these components need to rewrite or reformulate the input question. 
\end{inparaenum}
~\\
{\bf Impact of Abbreviations in Entities.} \begin{inparaenum}[\itshape i\upshape)]
The abbreviations presented in the question have an impact on the performance of the NED components. For the question \qq{What is the time zone of New York City?}, both TagMe and DBpedia Spotlight can disambiguate the entity \texttt{New York City} to \texttt{dbr:New\_York\_City}. However, when this question is rephrased as \qq{What is the time zone of NYC?}, TagMe succeeds while DBpedia Spotlight fails.
\end{inparaenum}
~\\
{\bf Impact of Number of Words in Entities.} \begin{inparaenum}[\itshape i\upshape)]
For the fourth question feature, regarding the question \qq{List all the chairmen of Vancouver Whitecaps FC?}, the segment \texttt{Vancouver Whitecaps FC} is expected to be linked to \texttt{dbr:Vancouver\_Whitecaps\_FC}.
Although the overall performance of DBpedia Spotlight is lower than TagMe, it succeeds in linking the entity while TagMe fails to recognise and disambiguate \texttt{Vancouver Whitecaps FC} to its target entity. 
In this question, the entity \texttt{Vancouver Whitecaps FC} has three words. We observe that if the number of words in the entity increases further, then DBpedia Spotlight also fails to identify entities. For instance, for the question \qq{Who has been a manager of english under twenty one football team?}, both components fail to disambiguate the entity \texttt{english under twenty one football team} (6 words) to its target DBpedia entity. 
\end{inparaenum}
~\\
These observations lead us to assume that certain characteristics of input questions such as length, variety of expressions and number of words in entities impact the performance of QA components.
In other words, a fine-grained analysis of performance indicates that the overall performance of a component is not representative when expressed as an average performance value on all the questions of a dataset because we observe a correlation between the input question characteristics and the success rate.
\section{Related Work} \label{sec:relatedwork}
~\\
{\bf QA Systems and Evaluation.} \begin{inparaenum}[\itshape i\upshape)]
Question answering systems for factoid questions over different publicly available Knowledge Graphs (KGs) have been evaluated using various datasets/benchmarks. SimpleQuestion~\cite{DBLP:journals/corr/BordesUCW15} and WebQuestions \cite{DBLP:conf/emnlp/BerantCFL13} are popular datasets for evaluating systems that use Freebase as underlying KG. Most of the state-of-the-art QA systems over Freebase have been evaluated using the SimpleQuestions and WebQuestions dataset~\cite{DBLP:conf/emnlp/BerantCFL13,DBLP:conf/acl/DaiLX16,DBLP:conf/acl/YuYHSXZ17}. For DBpedia, Question Answering over Linked Data (QALD) benchmark series provides widely used datasets for evaluating QA systems~\cite{DBLP:conf/clef/UngerFLNCCW15}. The number of questions in the QALD datasets ranges from 50 to 350 across its different editions. Therefore, state-of-the-art QA systems over DBpedia have been evaluated mostly on limited number of questions (50--100)~\cite{DBLP:journals/pvldb/CuiXWSHW17,GERBILQA,DBLP:conf/sigmod/ZouHWYHZ14}. LC-QuAD \cite{trivedi2017lc} is another recently released dataset that contains 5,000 questions for DBpedia and has been used for evaluating question answering systems~\cite{diefenbach2018towards} as well. 

Although the aforementioned benchmarks have been extensively used to evaluate QA systems as well as QA related components, no focus has been set on analysing the results with respect to the different types of questions included in these benchmarks. Therefore, across datasets and KGs, QA systems and components have reported their overall accuracy \textit{on average over all the questions present in the dataset} (i.e. by performing macro evaluation), which does not provide many insights about the weaknesses and strengths of a particular QA component and system when considering different types of questions.
\end{inparaenum}
~\\
{\bf Hybrid Question Answering System.} \begin{inparaenum}[\itshape i\upshape)]
Cui et al.~\cite{DBLP:journals/pvldb/CuiXWSHW17} introduced the concept of hybrid question answering systems in 2017. 
The authors combine their proposed KBQA QA system with existing monolithic QA systems to build hybrid QA systems. The evaluation shows that this increases the overall accuracy of the hybrid QA system for the 99 questions from QALD-3~\cite{cimiano2013multilingual} dataset. 
Singh et al.~\cite{DBLP:conf/www/SinghRBSLUVKP0V18} extend the applicability of hybrid QA systems to component level and combine different components per question answering task to build QA systems. The work has been evaluated using the 204 questions of QALD-5~\cite{DBLP:conf/clef/UngerFLNCCW15} and more than 3,000 questions of the LC-QuAD~\cite{trivedi2017lc} dataset. In these studies, the performance of hybrid QA systems and components has increased, however, the characteristics of the unanswered questions have not been further analysed. It is also not clear for which types of questions hybrid QA systems and components find limitations.
Therefore, in order to build hybrid QA systems collaboratively, it is also important to understand the strengths and weaknesses of different components that are used to build hybrid QA systems. 
\end{inparaenum}
\section{Evaluation Settings} \label{sec:setting}
In this section, we describe the evaluation settings of the experiments reported in this paper. In \autoref{sec:component}, we list the QA components considered in our experiments. Then, \autoref{sec:experimentalsetup} describes the experimental setup and employed metrics for assessment. \autoref{sec:benchmarking} explains the benchmarking procedure. Finally, \autoref{sec:feature} presents all features derived from the questions.

\subsection{Components for Evaluation} \label{sec:component}
\textcolor{black}{The \textit{Frankenstein framework}~\cite{DBLP:conf/www/SinghRBSLUVKP0V18} was employed as the underlying platform to run and evaluate QA components for our experiments since compared to the state of the art (i.e. openQA \cite{openqa}, QALL-ME \cite{qallme} and OKBQA \cite{kim2016open}) it includes a higher number of QA components. Frankenstein wraps the functionality of public APIs of these components and integrates them in its core QA pipeline as microservices. Therefore, it does not have any impact on the functionality of these components. However, most of the QA components when reused in QA frameworks target high precision rather than high recall per question. For instance, in the case of TagMe\footnote{\url{https://services.d4science.org/web/tagme/demo}} and for the given question \qq{Who is the wife of Barack Obama?} three entities with different confidence scores are returned (\texttt{Barack Obama}, \texttt{First\_Lady\_of\_the\_United\_States} and \texttt{World\_Health\_Organization}). Frankenstein picks entities with confidence score more than 0.50 which leads to the selection of only one entity i.e. Barack Obama (correct in case of this question to formulate SPARQL query), thus maximising the precision.}
Frankenstein currently integrates 29 QA components including 11 NER, 10 NED, five RL and two QB components that have been used to build QA systems\footnote{Full list of components: \url{https://github.com/WDAqua/Frankenstein/blob/master/Component\%20List.csv}}. \textcolor{black}{Except for 11 NER, rest of the NED components implicitly perform the NER task and provide directly the disambiguated URLs which are required to formulate the SPARQL query for an input question. Therefore, we focus on the NED task excluding the independent evaluation of NER components. However, all 11 NER components are used jointly with the AGDISTIS~\cite{UsbeckNRGCAB14} disambiguation component to evaluate how well these NER components recognise entities that can act as input for a disambiguation component such as AGDISTIS.}

Besides the most common QA tasks such as NED, RL and QB, in specific questions, it is necessary to map to the ontology classes\footnote{\url{https://www.w3.org/2002/07/owl\#Class}} present in the question. For the exemplary question \qq{Which comic characters are painted by Bill Finger?}, its intermediate representation using DBpedia (\qq{Which \texttt{dbo:ComicsCharacter} are \texttt{dbo:creator} by \texttt{dbr:Bill\_Finger}?}) requires recognising and linking to the expected DBpedia class \texttt{dbo:ComicsCharacter} as one of the inputs for constructing the corresponding SPARQL query. Frankenstein includes two Class Linking (CL) components. To the best of our knowledge, there are no other independent components (for DBpedia) except the 29 components integrated in Frankenstein for NED (20), RL (5), QB (2) and CL (2)\footnote{Based on the research published till 31/07/2018}.

\begin{figure*}[t]
	\centering
	\includegraphics[width=.9\textwidth]{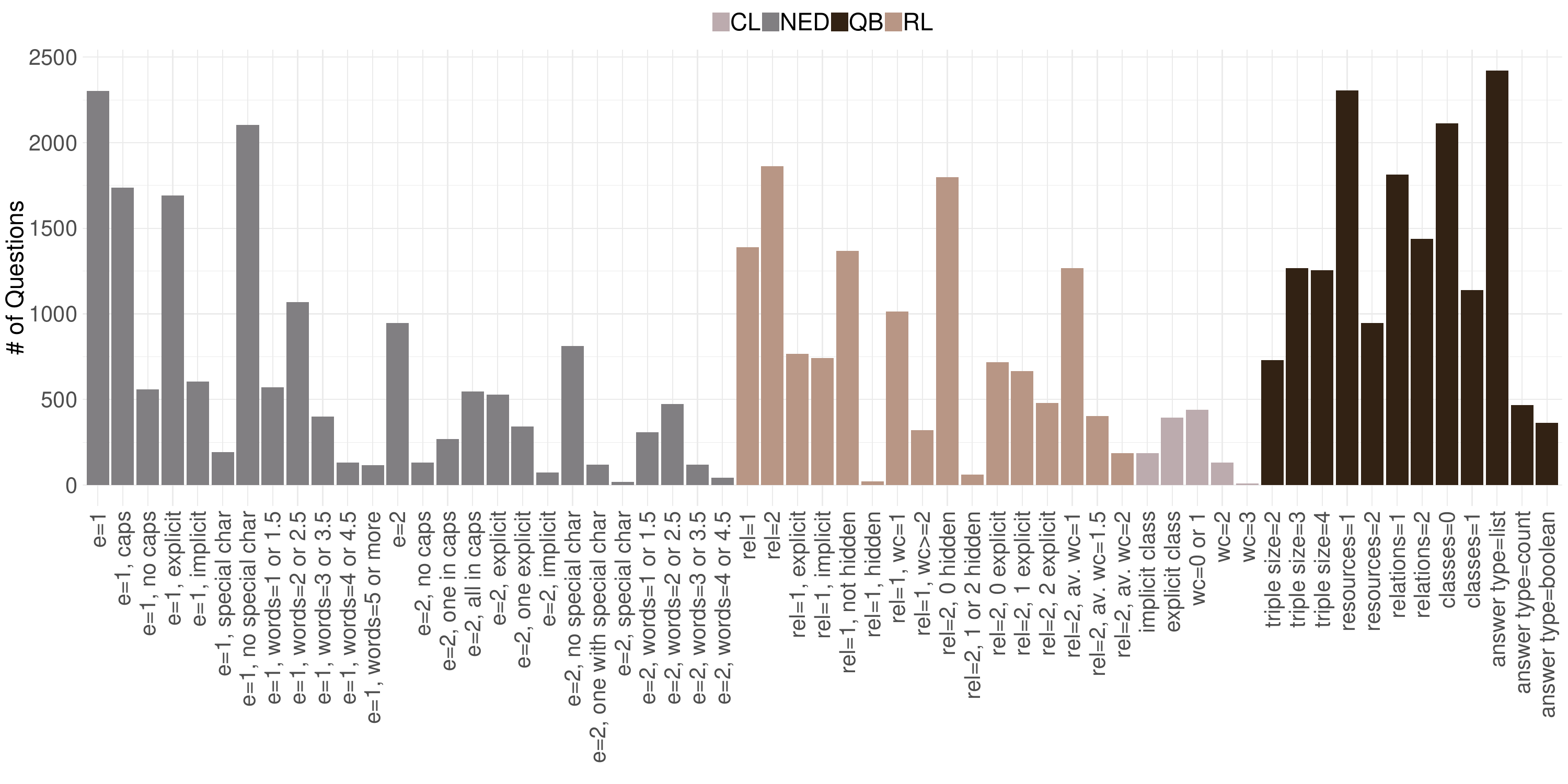}
	\caption{Distribution of Questions per Question Feature. For evaluating components fulfilling different QA tasks, various question features have been considered. 59 features have been extracted in total -- 26 for NED, 16 for RL, 5 for CL and 12 for QB. Some features are considered in several questions, and some questions are characterised by several features. 
	}
	\label{fig:feature-frequencies}
\end{figure*}

\subsection{Evaluation Setup} \label{sec:experimentalsetup}
We executed our experiments on ten virtual servers, each with eight cores, 32 GB RAM running on the Ubuntu 16.04.3 operating system. We have reused the open source implementation of Frankenstein\footnote{\url{https://github.com/WDAqua/Frankenstein}} for executing the different QA components and Stardog v4.1.3\footnote{\url{https://www.stardog.com/}} as RDF (Resource Description Framework)\footnote{\url{https://www.w3.org/RDF/}} datastore to store all data generated in the experiments.
~\\
{\bf Knowledge Graph.} \begin{inparaenum}[\itshape i\upshape)]
DBpedia\footnote{DBpedia version 2016-10} \cite{DBLP:conf/semweb/AuerBKLCI07} was our underlying background knowledge graph. It has 5.6 million entities and 111 million (subject,predicate,object) triples. It requires 14.2 GB storage.
\end{inparaenum}
~\\
{\bf Test Data.} \begin{inparaenum}[\itshape i\upshape)]
QALD\footnote{\url{https://qald.sebastianwalter.org/}} and LC-QuAD \cite{trivedi2017lc} are two datasets of \textit{factoid} questions available to benchmark QA systems and components over DBpedia. The total number of questions in the QALD series is relatively low (350 questions at most in the 5th edition) and the questions do not have much diversity.
Compared to QALD, the newly released LC-QuAD has 5,000 questions, the highest number of benchmarking questions available for DBpedia~\cite{trivedi2017lc}. The fully annotated LC-QuAD dataset (with entity, relation and class labels) has over 5,000 entities and more than 615 predicates covered in its questions. The questions are also diverse in expressiveness, as 20\% of questions are simple (questions with only one entity and one relation) and 80\% are complex questions. However, only 3,253 questions from this benchmark return answers from the latest DBpedia version\footnote{\url{http://dbpedia.org/sparql} (version 2016-10)} because the dataset was created using a previous DBpedia version 16-04. Since all components integrated in Frankenstein use the latest version of DBpedia for their functionalities, we have considered only this subset of LC-QuAD. These 3,253 questions were evaluated against the 29 Frankenstein components independently, resulting into 94,337 total questions executed using the Frankenstein framework. To the best of our knowledge, this is the first study over DBpedia with a high number of questions and components. Frankenstein platform generates a KG of results using
RDF (Resource Description Framework)\footnote{\url{https://www.w3.org/RDF/}}
format~\cite{DBLP:conf/esws/SinghBRS18} for further analysis. 
The execution of all questions over all available components generates approximately 4.9GB RDF data. The experimental data, the questions, source code for experiments and detailed results can be found in our open source repository for reusability and reproducibility\footnote{\url{HiddenforBlindReview}}.
\end{inparaenum}
~\\
\noindent
{\bf Metrics:} The following evaluation metrics have been employed: \begin{inparaenum}[\itshape i\upshape)]
\item \texttt{Micro Precision (MP)}: For a given component, the ratio of correct answers vs.\ total number of answers retrieved for a particular question.
\item \texttt{Precision}: The average of the Micro Precision over all questions by a component.
\item \texttt {Micro~Recall~(MR)}: The number of correct answers retrieved by a component vs.\ gold standard answers for the given question.
\item \texttt{Recall~(R)}: The average of Micro Recall over all questions for a given component.
\item \texttt{Micro~F-score~(MF)}: Harmonic mean of MP and MR for each question.
\item \texttt{F-score}: Harmonic mean of P and R for each component.
\item \texttt{Processed Question}: Question for which a component has non zero F-score.
\item \texttt{Unprocessed Question}: Question for which a component has zero Micro F-score.
\item \texttt{Baseline Value}: The highest F-score value for a particular question feature. 
\item \texttt{Baseline Component}: The component with the highest F-score value for a particular question feature.
\item \texttt{State-of-the-art F-score}: The highest F-score value per task on average over all questions.
\end{inparaenum}
\noindent

\subsection{Component Benchmarking} \label{sec:benchmarking}
In a particular QA task, the performance of each component is measured initially by calculating the Micro F-score per question, and then computing the Macro F-score representing the overall performance.
For instance, consider the question \qq{Which comic characters are painted by Bill Finger?} from the LC-QuAD dataset. The corresponding SPARQL query for this question is:
\begin{lstlisting}[language=SPARQL]
SELECT DISTINCT ?uri WHERE 
{?uri <http://dbpedia.org/ontology/creator> 
<http://dbpedia.org/resource/Bill_Finger> . ?uri 
<https://www.w3.org/1999/02/22-rdf-syntax-ns#type> 
<http://dbpedia.org/ontology/ComicsCharacter>}
\end{lstlisting}
In order to benchmark the NED components, the output URIs of the disambiguated entities produced by each component must be compared against the correct DBpedia entity (\texttt{dbr:Bill\:Finger} in this case). Afterwards, for each component Micro Precision, Micro Recall and Micro F-score values are calculated. A similar procedure has been followed for the RL and CL components, whose outputs are compared to the correct URIs (i.e. \texttt{dbo:creator} for the RL and \texttt{dbo:ComicsCharacter} for the CL task). For each of the QB components, given a correct set of URIs as input, the generated SPARQL query is compared to the benchmark SPARQL query for the given question by comparing the answers the SPARQL queries retrieve from DBpedia. A similar component benchmarking procedure has been followed in~\cite{DBLP:conf/i-semantics/MulangSO17,kcap}.
%
\subsection{Question Features} \label{sec:feature}
\textcolor{black}{Question classification based on features (e.g. question length, POS tags, head words etc.) has been a continuous field of research after the first edition of TREC challenge~\cite{voorhees2000overview} for open domain QA. However these features are explicitly used for answer type classification~\cite{DBLP:conf/emnlp/HuangTC09} and not for benchmarking/evaluating QA systems. In QA over KGs, NED, RL, CL and QB are concerned with the formation of final SPARQL queries and not with finding the right source of the answer. Therefore, for analysing the performance of QA components, and in order to provide a more fine-grained evaluation, we categorise questions per feature and evaluate the single QA components for these question features (cf.\ \autoref{fig:feature-frequencies} for the distribution of questions per question feature). All these question features have been automatically extracted.}

\subsubsection{Question Features for NED Task}
For the NED task, we consider the following features of input questions.
~\\
{\bf Number of Entities.} \begin{inparaenum}[\itshape i\upshape)]
For a given question, we consider the number of entities included in the question. For example, in the question \qq{Which comic characters are painted by Bill Finger?}, there is one entity (\texttt{Bill Finger}), and therefore this question is classified as ``question with entity=1''.
\end{inparaenum}
~\\
{\bf Number of Words in Entities.} \begin{inparaenum}[\itshape i\upshape)]
The number of words of an entity is the next feature. 
For the exemplary question \qq{Which comic characters are painted by Bill Finger?}, the number of words in the entity label (i.e. \texttt{Bill Finger}) is two; hence, this question is annotated with the feature ``entity=1, number of words in entities=2''. For questions having several entities, the average number of words in all labels of entities is considered.
\end{inparaenum}
~\\
{\bf Implicit/Explicit Entities.} \begin{inparaenum}[\itshape i\upshape)]
The entity of a question is explicit if there is an exact match between the segment and the DBpedia resource URI label (meaning there is no vocabulary mismatch). For the exemplary question, the segment \texttt{Bill Finger} matches the label of \texttt{dbr:Bill\_Finger}, while for the question \qq{What religions do politicians in the Korean Minjoo Party follow?}, the segment \texttt{Korean Minjoo Party} and the desire resource \texttt{dbr:Minjoo\_Party\_of\_Korea} have a slight vocabulary mismatch.
We consider the explicit/implicit entities as another feature. 
Please note that questions with two entities might have three different statuses 1) both entities are explicit, 2) at least one of them is explicit and 3) both are implicit.
\end{inparaenum}
~\\
{\bf Case Sensitivity of Entities.} \begin{inparaenum}[\itshape i\upshape)]
The use of uppercase or lowercase for entities is another question feature. If all words in the segment associated to an entity of a given question are uppercase, then we annotate the question as ``uppercase'', otherwise as ``lowercase''. For questions containing two entities, three different annotations for this feature are: 1) both entities are in uppercase, 2) at least one entity is in uppercase and 3) both entities are in lowercase.
\end{inparaenum}
~\\
{\bf Special Characters in Entities.} \begin{inparaenum}[\itshape i\upshape)]
If the entity segment contains ASCII punctuation, symbols or numbers, the question is annotated as a question with special character(s). For example, the question \qq{What is the route junction of Rhode Island Route 15?} contains a number in the entity segment \texttt{Rhode Island Route 15}, therefore it is considered as a question with special characters. For questions with two entities, three annotations are possible: 1) both entities have special characters, 2) at least one has special character and 3) none has special characters.
\end{inparaenum}

\subsubsection{Question Features for RL Task}
The features for the RL task are similar to the NED task except that we excluded (i) case sensitivity and (ii) inclusion of special characters, because there were no relations with such features in the LC-QuAD questions.
~\\
{\bf Number of Relations.} \begin{inparaenum}[\itshape i\upshape)]
The number of relations in a given question is the first question feature.
\end{inparaenum}
~\\
{\bf Number of Words in Relation.} \begin{inparaenum}[\itshape i\upshape)]
The number of words in the relation segment is the next feature. For questions with two relations or more, this feature is computed as the average number of words in the relation segments.
\end{inparaenum}
~\\
{\bf Explicit/Implicit Relation.} \begin{inparaenum}[\itshape i\upshape)]
Similar to the NED task, this feature refers to whether the relation mention is explicit or implicit. In the exemplary question \qq{Which comic characters are painted by Bill Finger?}, the relation segment \texttt{painted by} does not explicitly match to the DBpedia relation \texttt{dbo:creator}, therefore, this question is considered as implicit relation question.
\end{inparaenum}
~\\
{\bf Hidden Relation.} \begin{inparaenum}[\itshape i\upshape)]
Consider the question \qq{Which companies have launched a rocket from Cape Canaveral Air Force station?} which contains two relations (i.e. \texttt{dbo:launchSite} and \texttt{dbo:manufacturer}). For \texttt{dbo:launchSite}, there is a mention in the question (\texttt{launched}), while, there is no mention for the relation \texttt{dbo:manufacturer}. We characterise questions without a natural language segment as ``questions with hidden relations''. For questions with two relations, possible annotations are: 1) one relation is hidden and 2) at least one is hidden.
\end{inparaenum}

\subsubsection{Question Features for CL Task}
All LC-QuAD questions contain at most one class. For the questions with a single class, we considered the following features: 1) explicit/implicit mention of class and 2) number of words in the class segment. 

\subsubsection{Question Features for QB Task}
For the evaluation of the QB components, we take into consideration the following features: 
~\\
{\bf Number of Triple Patterns.} \begin{inparaenum}[\itshape i\upshape)]
This feature refers to the number of basic triple patterns in the SPARQL query corresponding to the question.
\end{inparaenum}
~\\
{\bf Number of Relations.} \begin{inparaenum}[\itshape i\upshape)]
It means the number of relations in the SPARQL query.
\end{inparaenum}
~\\
{\bf Number of Entities.} \begin{inparaenum}[\itshape i\upshape)]
It relates to the number of entities in the SPARQL query.
\end{inparaenum}
~\\
{\bf Number of Classes.} \begin{inparaenum}[\itshape i\upshape)]
This feature refers to the number of classes in the SPARQL query.
\end{inparaenum}
~\\
{\bf Answer Type.} \begin{inparaenum}[\itshape i\upshape)]
There are three types of possible types: list, number or boolean. For instance, the question \qq{Which comic characters are painted by Bill Finger} has a list as expected answer.
\end{inparaenum}
\section{Evaluation Results} \label{sec:results}

We pursue the following research questions in our experiments.
\begin{inparaenum}[\bf {\bf RQ}1\upshape)]
    \item What is the performance of QA components depending on the question features?
    \item Which QA components exhibit similar behaviour across the different questions and question features?
    \item What is the impact of combining QA components in the overall performance?
\end{inparaenum}
This section is followed by the results of the empirical experiments conducted to address our research questions. \autoref{sec:component-performance} represents the experiments with respect to RQ1.
In \autoref{sec:clustering} and \autoref{sec:hybrid} we discuss our observations and answers reached for RQ2 and RQ3 respectively.



\subsection{Evaluating QA Components} \label{sec:component-performance}
In this experiment, we evaluate 29 components for various QA tasks based on the question features described in \autoref{sec:feature}. We then report the results for each QA task independently.

\begin{figure}[t]
	\centering
	\includegraphics[width=\columnwidth]{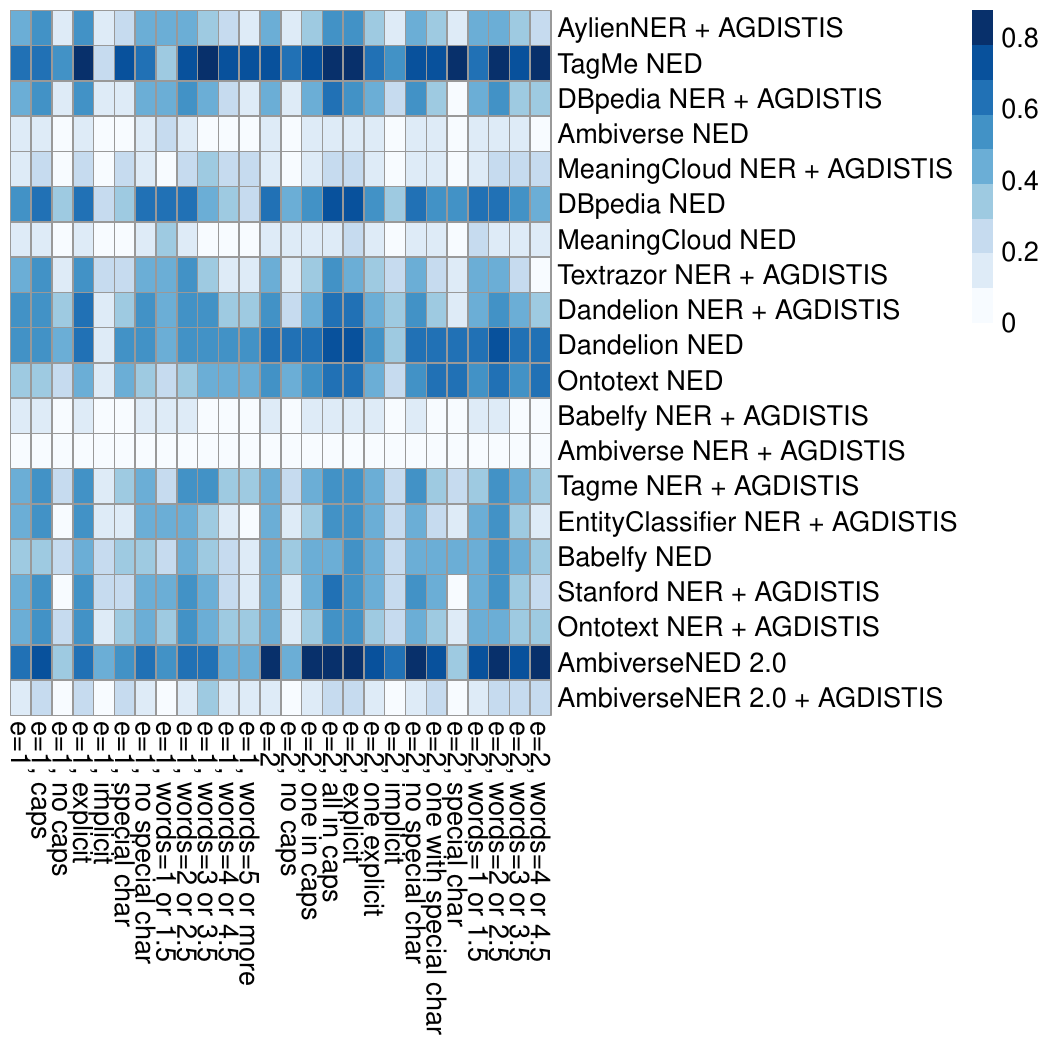}
	\caption{
	Comparison of F-scores per Question Feature for QA Components Performing the NED Task. Darker colour indicates higher F-score. While TagMe and AmbiverseNED 2.0 demonstrate high performance, few components such as Babelfy, Ambiverse 1.0 and MeaningCloud deliver poor results for the LC-QuAD questions. These results suggest that the behaviour of best performing NED components is impacted by the question features.}
	\label{fig:heatmap-ned}
\end{figure}

\subsubsection{Performance of NED Components} \label{sec:nedc}
The heatmap in \autoref{fig:heatmap-ned} shows the performance (i.e. F-score) for 20 NED components across questions with different features.
For example, TagMe~\cite{DBLP:conf/cikm/FerraginaS10} which in previous literature reported F-score value 0.67~\cite{DBLP:conf/www/SinghRBSLUVKP0V18}, has fluctuating F-scores here.
~\\
{\bf Performance varies across question features.} \begin{inparaenum}[\itshape i\upshape)]
The baseline F-score value varies across different question features. For example, for the question feature ``number of entities=1, and contains special character'' (e.g. question \qq{What is the route junction of Rhode Island Route 15?} where the entity label \texttt{Rhode Island Route 15} contains special characters), TagMe~\cite{DBLP:conf/cikm/FerraginaS10} is the baseline component with F-score value of 0.72. However, for questions such as \qq{What religions do politicians in the Korean Minjoo Party follow?}, in which the number of entities is one and the entity is implicit (the segment \texttt{Korean Minjoo Party} is mapped to \texttt{dbr:Minjoo\_Party\_of\_Korea}), AmbiverseNED 2.0\footnote{\url{https://www.ambiverse.com/}}~\cite{DBLP:conf/emnlp/HoffartYBFPSTTW11} is the baseline component, but with lower baseline value -- 0.41.
\end{inparaenum}
~\\
{\bf Impact of Number of Entities.} \begin{inparaenum}[\itshape i\upshape)]
When the number of entities in the question is two (e=2), 16 NED components illustrated an increase in the F-score values compared to questions with one entity. For such questions, AmbiverseNED 2.0 is the baseline component with F-score 0.79. When the number of entities is one, all components have F-scores less than 0.67 which is the state-of-the-art F-score for the NED task. In this case, TagMe sets the baseline F-score to 0.65.
%
\end{inparaenum}
~\\
{\bf Impact of Number of Words in Entities.} \begin{inparaenum}[\itshape i\upshape)]
For questions which contain one entity, if the number of words ("wc" in heatmap) in the entity increases from one to two, 19 out of the 20 NED components demonstrate an increase in the F-score values. However, when the number of words in the entity further increases to three, 16 components demonstrate decrease in F-scores. For four or more words in the entity, only OntotextNED\footnote{\url{https://ontotext.com/technology-solutions/semantic-tagging/}} shows no variation in the F-score compared to questions with entities containing three words; for all other components, F-score is further decreased.
In questions containing two entities, the highest F-score value remains above 0.75 independently of the average number of words in the entities. 
\end{inparaenum}
~\\
{\bf Impact of Special Characters in Entities.} \begin{inparaenum}[\itshape i\upshape)]
Except for TagMe and DandelionNED\footnote{\url{https://dandelion.eu/docs/api/datatxt/nex/getting-started/}}, for all other NED components we observed a decrease in the F-score values in the case that the spotted entities contain special characters.
\end{inparaenum}
~\\
{\bf Impact of Character Cases in Entities.} \begin{inparaenum}[\itshape i\upshape)]
The character cases significantly impact the performance of the NED components. The majority of QA components demonstrate a steep decrease in the F-score values when the entities appear in lowercase, independent of the number of entities in the question. For example, in DBpedia Spotlight NED \cite{MendesJGB11} (abbreviated as DBpedia NED in \autoref{fig:heatmap-ned}) the F-score decreases from 0.64 to 0.36 when the entity character case is switched to lowercase. Similar behaviour is observed in other components as well such as AylienNER+AGDISTIS\footnote{\url{http://docs.aylien.com/docs/introduction} (AylienNER)} \cite{UsbeckNRGCAB14} and StanfordNER+AGDISTIS~\cite{DBLP:conf/acl/FinkelGM05,UsbeckNRGCAB14}, for which the F-score values sharply drop from 0.50 and 0.55 to 0.15 and 0.09 respectively.
\end{inparaenum}
~\\
{\bf Impact of Implicit/Explicit Entities.} \begin{inparaenum}[\itshape i\upshape)]
The existence of implicit entities significantly impacts the performance of all NED components. For example, for questions with one explicit entity, the baseline F-score value is 0.78 (TagMe), whereas for implicit entities, the baseline is 0.41 (AmbiverseNED 2.0) and for the other 19 NED components is less than 0.30.
\end{inparaenum}
\subsubsection{Performance of RL Components}
\label{sec:rlc}
The heatmap in \autoref{fig:heatmap-rl} shows the performance (i.e. F-score) for five RL components across questions with different features.
The state-of-the-art F-score value for the RL task on LC-QuAD dataset is 0.23, achieved by the RNLIWOD component~\cite{DBLP:conf/www/SinghRBSLUVKP0V18}. Similar to the results reported for the NED task, there is no overall baseline component and value for all question features, as illustrated in \autoref{fig:heatmap-rl}. This experiment led us to the following observations: 
\begin{figure}[b]
	\centering
	\includegraphics[width=\columnwidth]{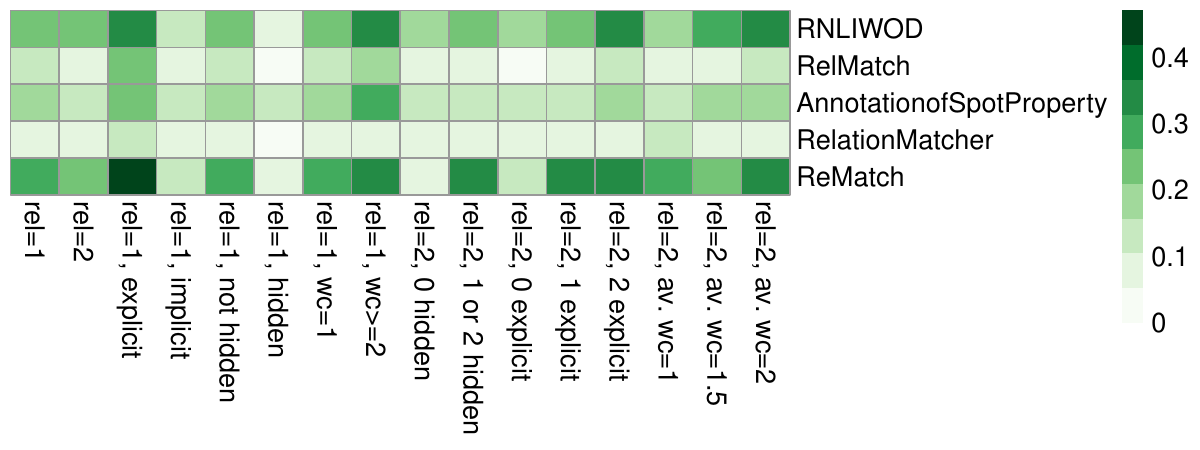}
	\caption{
	Comparison of F-scores per Question Feature for QA Components Performing the RL Task. Darker colour indicates higher F-score. We observe that ReMatch and RNLIWOD deliver the best F-scores but reach zero F-score for some question features. These results provide evidence that the performance of RL components is affected by most of the question features.}
	\label{fig:heatmap-rl}
\end{figure}
~\\
{\bf Impact of Number of Relations.} \begin{inparaenum}[\itshape i\upshape)]
When the number of relations (rel=1 or 2 in \autoref{fig:heatmap-rl}) in the question is elevated to two, the baseline F-score drops from 0.28 (ReMatch~\cite{DBLP:conf/i-semantics/MulangSO17}) to 0.24 (RNLIWOD\footnote{The component is similar to the Relation Linker of \url{https://github.com/dice-group/NLIWOD}} component). Other components also demonstrate a decrease in performance in this case.
\end{inparaenum}
~\\
{\bf Impact of Number of Words in Relations.} \begin{inparaenum}[\itshape i\upshape)]
For questions containing one or two relations, an increase in the number of words in the relation label has increased the overall performance of the components. 
\end{inparaenum}
~\\
{\bf Impact of Explicit/Implicit Relations.} \begin{inparaenum}[\itshape i\upshape)]
For questions containing one relation, for which the label is explicit, ReMatch~\cite{DBLP:conf/i-semantics/MulangSO17} provides the baseline F-score 0.47. However, when the relation is implicit, the baseline F-score value drops to 0.16 (for RNLIWOD). RelMatch~\cite{kim2016open} reports the lowest F-score for implicit relations with F-score 0.06.
In questions containing two relations, the number of implicit relations negatively impacts the performance. For instance, for questions being similar to \qq{How many currencies are in use in places where people speak French?} which has two implicit relations (i.e. \texttt{currencies} and \texttt{speak}) that should be mapped to \texttt{dbo:currency} and \texttt{officialLanguage} respectively, the baseline F-score value is 0.17 (for RNLIWOD).
\end{inparaenum}
~\\
{\bf Impact of Hidden Relations.} \begin{inparaenum}[\itshape i\upshape)]
For questions containing a natural language label for the relation, ReMatch is the baseline with F-score 0.28. However, for the exemplary question \qq{How many shows does HBO have?} There is no corresponding segment in the annotated dataset to be mapped to \texttt{dbo:company}. For all questions with one hidden relation, the baseline F-score is drops to 0.15. A similar drop in performance is observed in questions with two relations as well.
\end{inparaenum}
\begin{figure}[thp]
	\centering
	\includegraphics[width=.45\columnwidth]{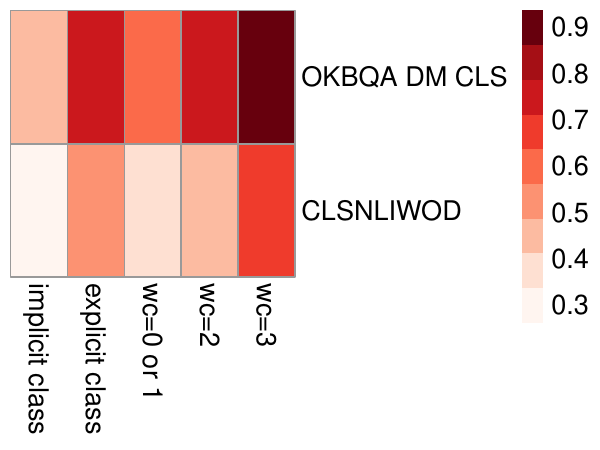}
	\caption{
	Comparison of F-scores per Question Feature for QA Components Performing the CL Task. Higher F-score is indicated with dark colour. OKBQA DM Class Linker performs better than CLSNLIWOD. However, all question features impact the performance of both components.
	}
	\label{fig:heatmap-cl}
\end{figure}

\subsubsection{Performance of CL Components} \label{sec:clc}
The heatmap in \autoref{fig:heatmap-cl} shows the fine-grained results of our experiment for the CL task. For this, we took into account two state-of-the-art components integrated in Frankenstein. We concluded that the following question features impact their overall performance:
~\\
{\bf Impact of Explicit/Implicit Class.} \begin{inparaenum}[\itshape i\upshape)]
Consider the question \qq{What university campuses are situated in Indiana?} for which the class label \texttt{university} is expected to be mapped to \texttt{dbo:University} through an exact match. In such cases, OKBQA DM CLS\footnote{Component is similar to the CL of \url{http://repository.okbqa.org/components/7}} has the baseline F-score 0.72 compared to the state-of-the-art F-score 0.52 for the CL task~\cite{DBLP:conf/www/SinghRBSLUVKP0V18}. 
However, when class mentions are implicit, the baseline F-score drops to 0.45, but OKBQA DM CLS is still the best component. 
\end{inparaenum}
~\\
{\bf Impact of Number of Words in Class.} \begin{inparaenum}[\itshape i\upshape)]
The increasing number of words in the class label has a positive impact on the performance of the components. OKBQA DM CLS remains the baseline for all question features. However, the baseline value increases with the number of words in the class label. When the class label contains one word the baseline F-score is 0.61; for two and three words, the baseline F-score increases to 0.72 and 0.94 respectively.
\end{inparaenum}
%
%
\subsubsection{Performance of QB Components} \label{sec:qbc}
The two QB components available in Frankenstein were evaluated based on different question features.
The details of performance behaviour are illustrated in \autoref{fig:heatmap-qb}. The state-of-the-art F-score for the QB task is 0.48 for NLIWOD QB\footnote{Component is based on \url{https://github.com/dice-group/NLIWOD}.}. We noticed the following observations based on the impact of the question features on the performance:
~\\
{\bf Impact of Number of Triple Patterns.} \begin{inparaenum}[\itshape i\upshape)]
With the increase in the number of triple patterns, the performance of the QB components is downgraded in general. For queries with two triple patterns, SINA~\cite{DBLP:journals/ws/ShekarpourMNA15} is the baseline component with F-score value 0.80. When the number of triple patterns is three, SINA's F-score drops to 0.18, and NLIWOD QB is the new baseline component with F-score 0.52. In the case of questions mapped to SPARQL queries with four triple patterns, NLIWOD QB remains the baseline with F-score 0.38, while the F-score for SINA drops to 0.
\begin{figure}[b]
	\centering
	\includegraphics[width=.75\columnwidth]{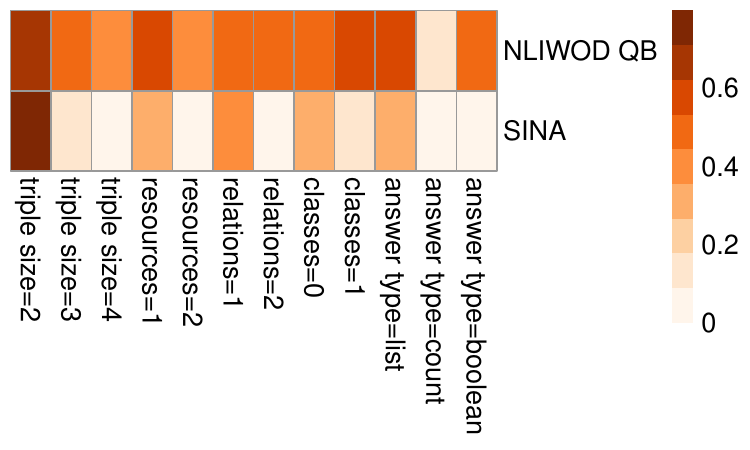}
	\caption{
	Comparison of F-scores per Question Feature for QA Components Performing the QB Task. Higher F-score is indicated with dark colour. SINA performs better than NLIWOD Query Builder only when the two triples are included in the input question. For most of the question features F-scores remain lower than 0.6.
	}
	\label{fig:heatmap-qb}
\end{figure}
\end{inparaenum}
\begin{figure*}[hpbt]
\centering
  \subfloat[Clustering based on question features (macro level)]{
      \includegraphics[width=0.49\textwidth]{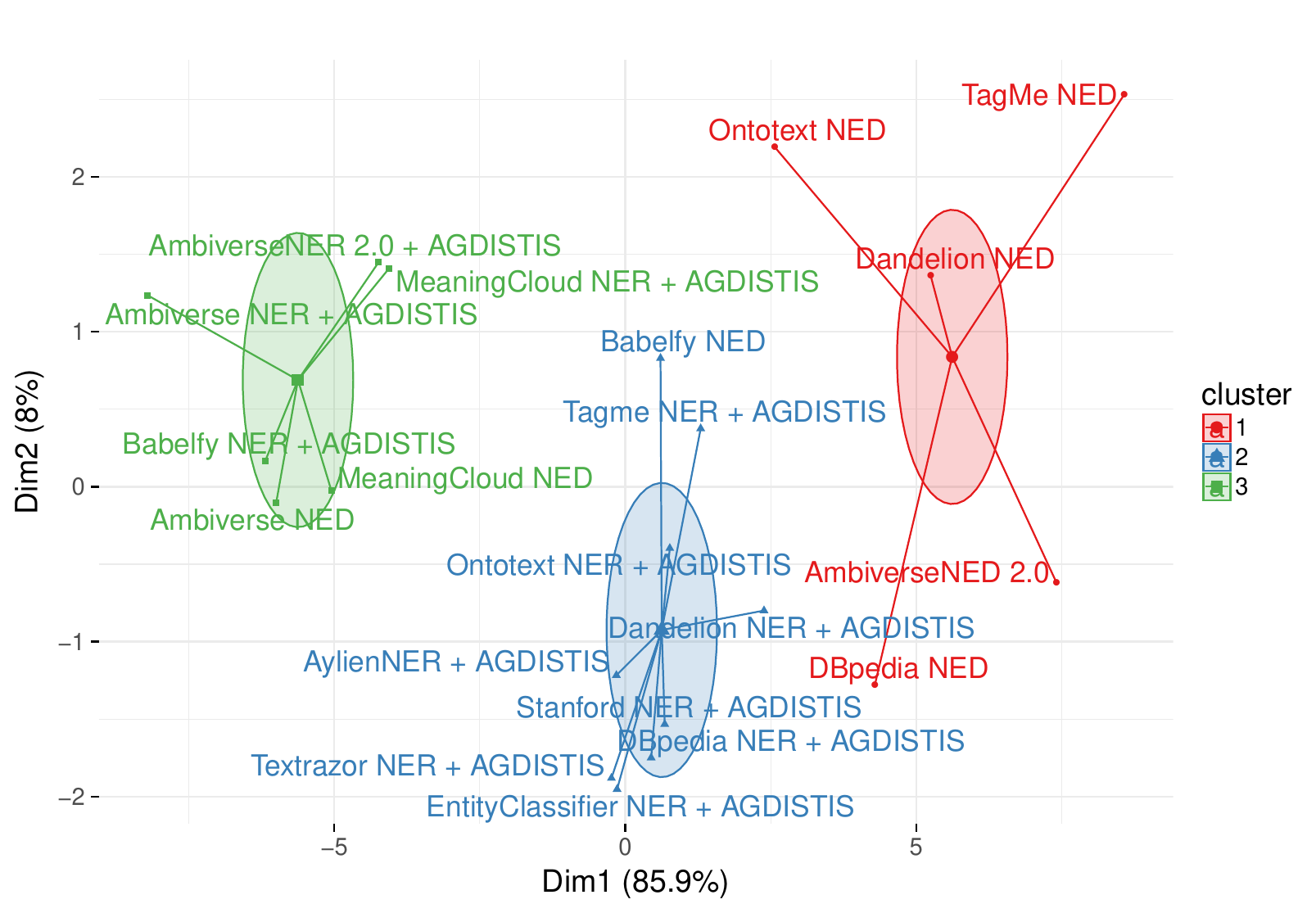}
      \label{fig:ned-clustering-1}}
  \subfloat[Clustering based on question results (micro level)]{
      \includegraphics[width=0.49\textwidth]{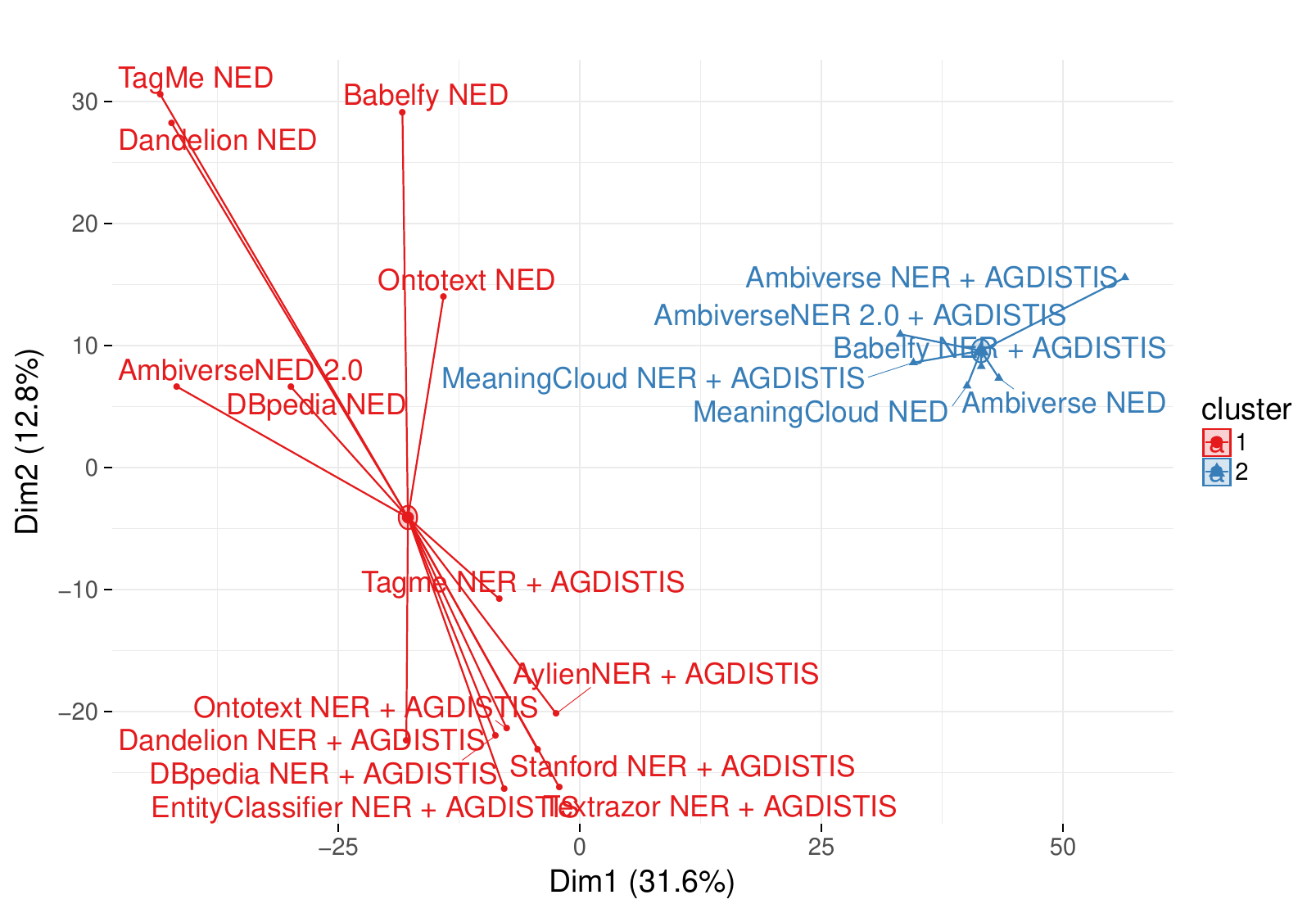}
      \label{fig:ned-clustering-2}}
    \caption{
   Clustering of NED Components. K-means clusters NED components at macro and micro level. In (a) the red cluster contains the best performing components while the black and the green clusters represent NED components with medium and low performance respectively, for questions with different features. While the components with similar performance are clustered together in (a), in (b) we observe that some are dissimilar with regard to the questions they are able to answer. 
    }
    \label{fig:ned-clustering}
\end{figure*}
~\\
{\bf Impact of Number of Resources.} \begin{inparaenum}[\itshape i\upshape)]
The increase in the number of resources (i.e. entities) in the question also inversely affects the performance of the QB components. In case of single resources, NLIWOD QB has F-score 0.54, while this value drops to 0.38 when the number of resources in the question increases to two. The F-score for SINA drops to 0 for questions with two resources.
\end{inparaenum}
~\\
{\bf Impact of Number of Relations.} \begin{inparaenum}[\itshape i\upshape)]
NLIWOD QB and SINA have shown dissimilar behaviour in the performance when the number of relations in the question increases. For an increase in the number of relations from one to two, the F-score value for NLIWOD QB has been elevated from 0.48 to 0.50, whereas the F-score value for SINA has dropped from 0.40 to 0.04.
\end{inparaenum}
~\\
{\bf Impact of Number of Classes.} \begin{inparaenum}[\itshape i\upshape)]
When there is no class in the question, NLIWOD QB is the baseline component (F-score is 0.46). For questions with one class, the baseline F-score value increases to 0.54 for NLIWOD QB which outperforms SINA in both cases.
\end{inparaenum}
~\\
{\bf Impact of Answer Type.} \begin{inparaenum}[\itshape i\upshape)]
NLIWOD QB remains the best component for any answer type (i.e. list, boolean and number.)
For instance, the question \qq{Which comic characters painted by Bill Finger?} expects a list of DBpedia resources as an answer. For such questions, NLIWOD QB is the baseline component with F-score 0.56. When the expected answer type is boolean, the baseline F-score value drops to 0.46, and it further drops to 0.14 for questions expecting a number as an answer. SINA cannot answer any questions with answer types number or boolean.
\end{inparaenum}

\subsection{Clustering of QA Components} \label{sec:clustering}
In this experiment, we pursue the second research question, that is where or when QA components exhibit similar behaviour. To study the similarity of QA components behaviour we applied clustering techniques \textcolor{black}{on the evaluation results of these components based on the question features as well as overall question results}. In particular, we have applied k-means clustering\footnote{The optimal k was calculated using the elbow and silhouette method \cite{Rousseeuw:1987:SGA:38768.38772}.} 
To cluster QA components based on (a) performance with respect to question features (macro level) and (b) performance with respect to processed LC-QuAD questions (micro level).
Figure~\ref{fig:ned-clustering-1} and Figure~\ref{fig:ned-clustering-2} visualise our clustering results. Since the number of components per QA task under evaluation differs, we were able to study this research question only for the task for which there is an adequate number of components, i.e. the NED task.
In the first clustering (cf.\ Figure~\ref{fig:ned-clustering-1}), the red cluster (right) represents components with higher performance, while the black (middle) and green (left) clusters group medium and low accuracy components respectively. \textcolor{black}{The ``distance'' between the single components can also be observed by comparing the colour scales in \autoref{fig:heatmap-ned}.} In the second clustering (cf.\ Figure~\ref{fig:ned-clustering-2}), components that appear in the same cluster (or are ``closer'' in the cluster) are expected to provide correct answers to similar subsets of the LC-QuAD questions. 
By comparing these two figures, we observe that NED components that are similar with respect to their performance over different question features are not necessarily able to answer the same questions as well \textcolor{black}{(although a tendency is in general observed)}. For instance, TagMe NED and Babelfy NED \cite{DBLP:journals/tacl/0001RN14} which appear close in the clustering of Figure~\ref{fig:ned-clustering-2} demonstrate different performance in Figure~\ref{fig:ned-clustering-1}. In addition, in order to find out which question features are more significant for clustering NED components, we have performed Principal Component Analysis (PCA). From this, we concluded that ``e=2, explicit'', ``e=2, words=2 or 2.5'' and ``e=2, no special characters'' constitute the most important features. This means, in principle, that the NED components under study demonstrate very different F-scores for these features. \textcolor{black}{This type of analysis allows us to group components with respect to the type of questions they are able to address successfully.}

\subsection{Hybrid Composition of QA Components} \label{sec:hybrid}
\begin{figure}[t]
	\centering
	\includegraphics[width=\columnwidth]{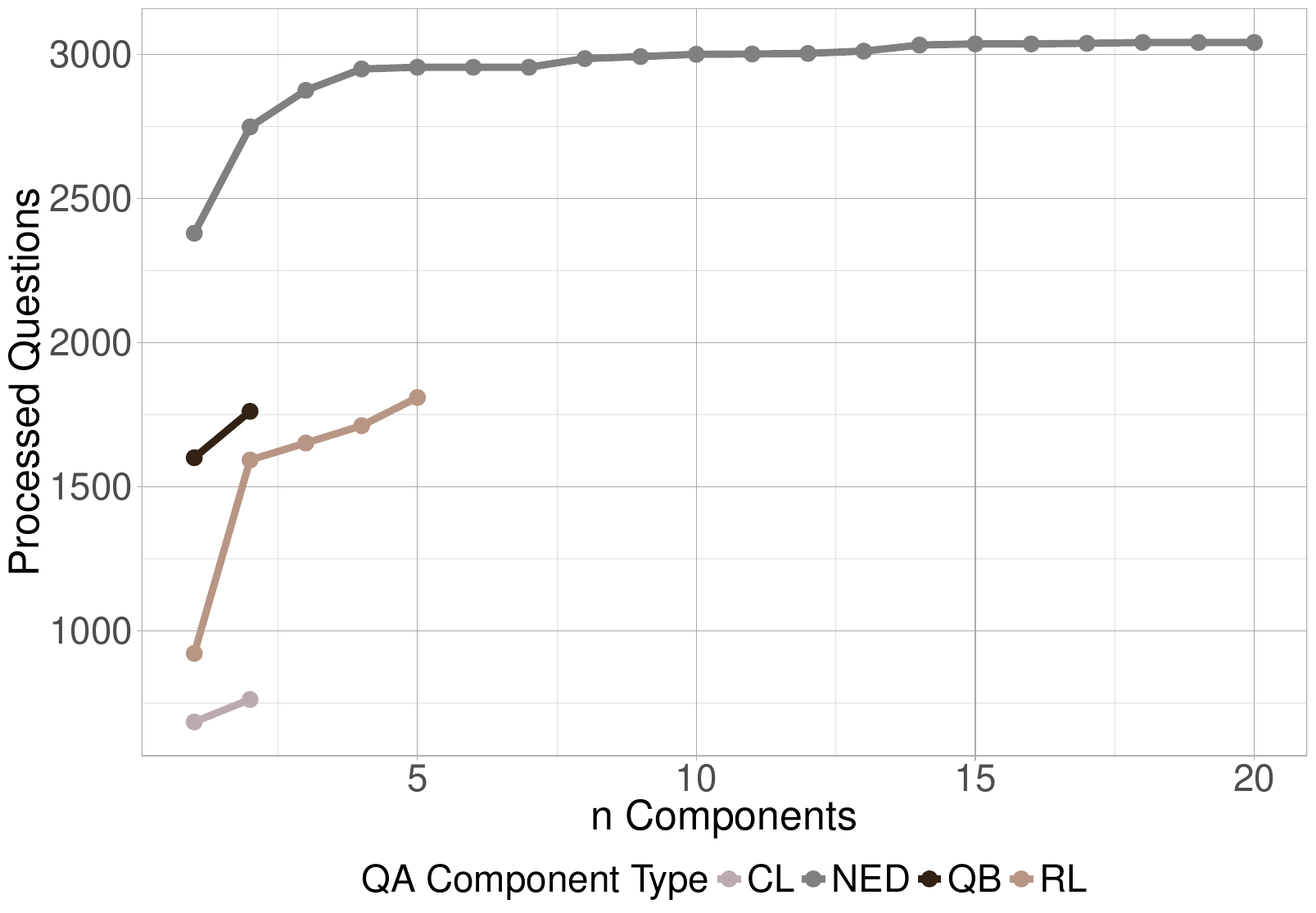}
	\caption{
	Total Processed Questions for the Combination of Top-n QA Components per QA Task. Combinations of two or more QA components improve the number of processed questions. Particularly, combinations of the top-4 NED are able to process a bit less than 3,000 questions; however, adding more NED components does not improve the performance significantly.}
	\label{fig:top-components}
\end{figure}

In this section, we address the third research question, in particular, we investigate what is the impact of combining components in the total performance.
The idea of integrating QA systems was first introduced by Cui et al.~\cite{DBLP:journals/pvldb/CuiXWSHW17}, who combined two QA systems and observed a slight increase in the number of answered questions for 99 questions of the QALD-3 dataset. First, the input question is sent to the KBQA QA system (introduced by the authors) and if KBQA cannot answer the question this question is passed to other monolithic QA systems; the combination is known as hybrid QA system. 
In this experiment, we extend the concept of hybrid QA systems by \textit{integrating at the component level} in order to study the impact of combining QA components on the total number of processed questions. We also use a bigger set of questions (3,253) from LC-QuAD compared to 99 questions from QALD used in \cite{DBLP:journals/pvldb/CuiXWSHW17}. For our experiment, we send a question to the component with highest F-score on LC-QuAD (average on all 3,253 questions), and if the question is not answered, we forward the question to the component with the next highest F-score. We continue this process for all QA tasks until the component with the lowest F-score.
We calculate the accumulative number of processed questions for top-1, top-2, etc.\ components which are taken into account.
\autoref{fig:top-components} illustrates the impact of combining QA components on the number of processed questions for all tasks. It can be observed that for the RL, QB and CL tasks this number continuously increases when more components are combined. However, for the NED task (where a bigger number of components is considered), the number of answered questions initially increases but it slowly saturates when more than four components are considered. This shows that combining several QA components performing the same task in a hybrid QA system does not necessarily improve the overall performance (not all the components are complementary). In addition, this observation provides an indication that most of the QA components fail for the same questions, i.e. there are several questions that cannot be addressed by any of the available components.

\begin{figure}[t]
	\centering
	\includegraphics[width=.72\columnwidth]{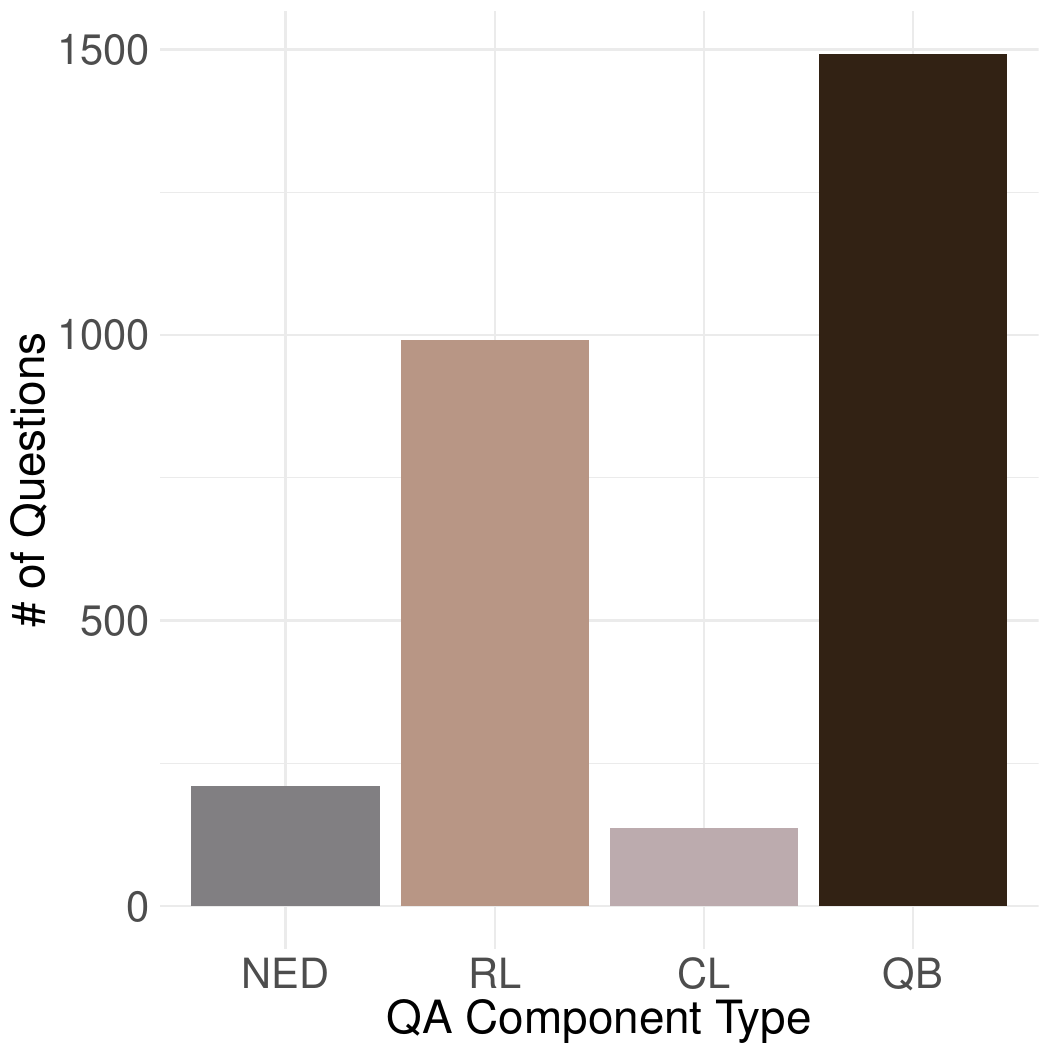}
	\caption{
	Number of Unprocessed Questions per QA Task. While combination of all NED and CL components are able to process the high majority of the LC-QuAD questions, the combinations of QB and RL components fails in half and one third of the questions, respectively.
	}
	\label{fig:totalunprocess}
\end{figure}

\begin{figure}[ht!]
	\centering
	\includegraphics[width=\columnwidth]{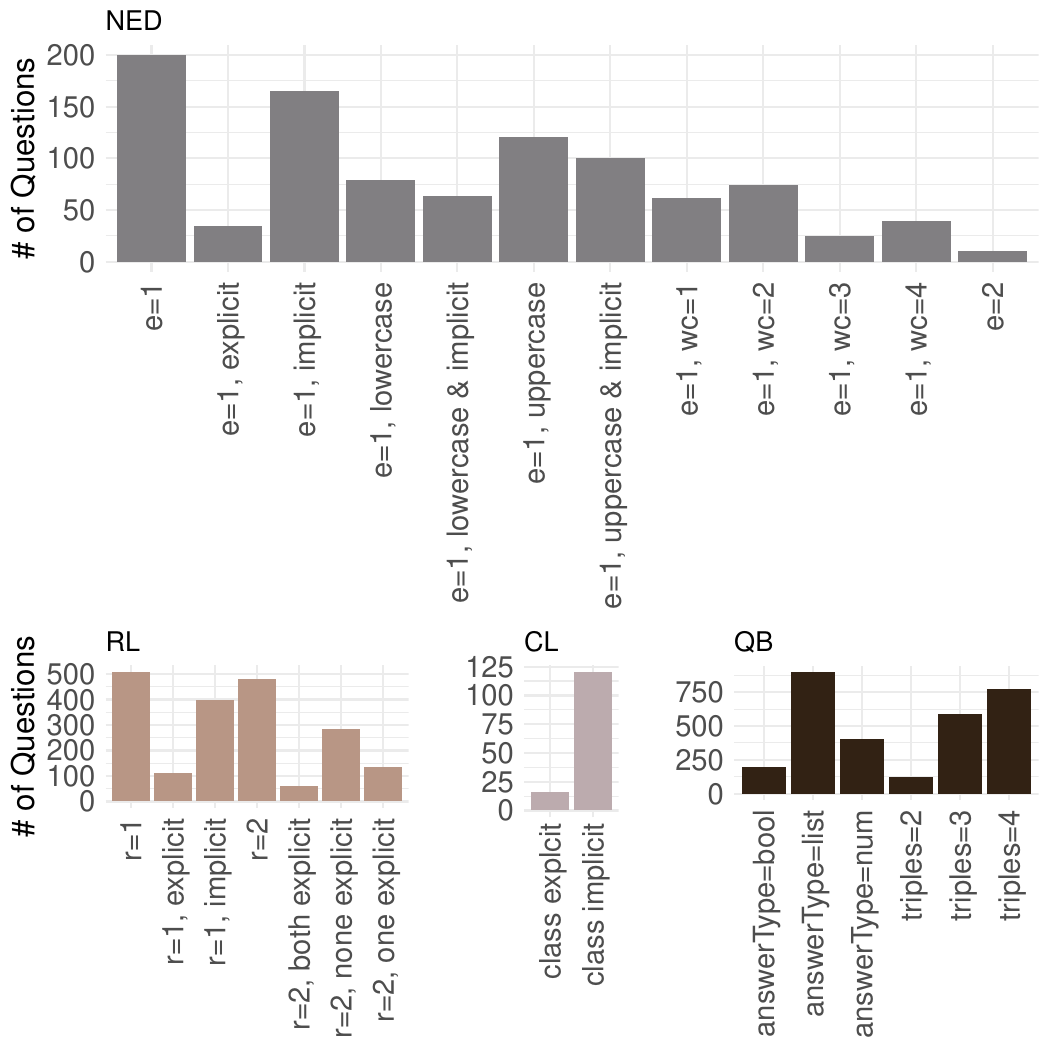}
	\caption{
	Distribution of Unprocessed Questions per QA Task Based on Question Feature. When the target entity, class or relation are implicit the corresponding QA components fail in many cases. For the QB components, generating SPARQL queries from questions with three or more triple patterns or expecting a number as an answer is at most challenging. This provides an evidence that few features majorly impact the performance of the components.
	}
	\label{fig:unprocessedtype}
\end{figure}

\section{Failure Analysis} \label{sec:failure}
There are several questions for which all components of the respective task failed to process (cf.\ \autoref{fig:totalunprocess}). We have analysed the question features for those questions per QA task in order to understand in depth the collective failure of the components. 
~\\
{\bf Unprocessed Questions for NED Components.} \begin{inparaenum}[\itshape i\upshape)]
For the NED task, 210 questions cannot be processed by any of the 20 components, i.e. for which the F-score for all components is 0. As illustrated in \autoref{fig:unprocessedtype}, 200 unprocessed questions have a single entity and just 10 questions have two entities. Out of the 200 single entity questions, 165 contain implicit entities. Furthermore, 121 questions with single entity have entity label written in uppercase letters and 79 with lowercase letters. While analysing these questions further, we have observed that out of 121 questions which have uppercase letters, 100 include implicit entities as well. A similar observation is valid for the 79 questions having an entity label in lowercase, out of which 64 contain an implicit entity. Hence, the implicit nature of entities is difficult to address by the majority of the NED components.
\end{inparaenum}
~\\
{\bf Unprocessed Questions for RL Components.} \begin{inparaenum}[\itshape i\upshape)]
Compared to the NED task, more questions for RL remain unprocessed. There are 992 questions in total for which all RL components have zero F-score. Among these questions, 510 questions have one and 482 questions have two relations. 397 relations are implicit when the number of relations is one. For the 482 questions with two relations, 135 questions have at least one implicit relation, whereas for 286 questions both relations are implicit.
\end{inparaenum}
~\\
{\bf Unprocessed Questions for CL Components.} \begin{inparaenum}[\itshape i\upshape)]
In total, 136 questions are not processed collectively by both CL components. These questions have predominantly implicit classes (in 120 cases) as can be seen in \autoref{fig:unprocessedtype}.
\end{inparaenum}
~\\
{\bf Unprocessed Questions for QB Components.} \begin{inparaenum}[\itshape i\upshape)]
The number of unprocessed questions for the two QB components is the highest compared to the other QA tasks i.e. 1,492 questions in total. The number of triple patterns in the SPARQL query is one of the major features observed in the unprocessed questions. Only 125 questions contain two triple patterns, while 591 questions have three and 776 questions four. Besides the number of triple patterns, it is obvious that answer type challenges the QB components. 195 unprocessed questions have answer type boolean, 402 number and 895 question expect a list as an answer. For 895 questions with list answer type, 839 have more than two triple patterns in the corresponding SPARQL query. 
\end{inparaenum}


\section{Discussion and Future Directions} \label{sec:discussion}
In the following paragraphs, we summarise our observations followed by a discussion about potential improvements in the state-of-the-art QA components.
~\\
\textbf{NED components with similar F-scores cannot always answer the same questions.} By exploring how ``close'' NED components are with respect to their F-scores for different question features and with regard to which questions they are able to answer, we observed that components that have similar performance on different question features may not be able to answer the same questions as well.
~\\
\textbf{Hybrid QA systems to some extent promotes the performance.}
Our experiment in combining QA systems at component level shows that the hybrid composition of QA components can increase the number of total processed questions (but fails in similar types of questions). However, specifically for NED components, we observed that with the increase of QA components the number of processed questions in total saturates.
~\\
{\bf Suggestions to improve NED Components.} \begin{inparaenum}[\itshape i\upshape)] 
\textcolor{black}{In 2014, Derczynski et al.~\cite{DBLP:journals/ipm/DerczynskiM0EGTPB15} pointed out capitalisation of entity label as an issue for NED tools (the authors analysed three tools) while analysing a tweet corpus. However, this issue still remains unsolved, and we identify sensitivity to character cases as major pitfall for 20 NED tools. We observe a decrease in performance for lowercase characters for all components, specifically, for 13 components, the performance drops more than 50\%. We are aware that many of NED components were not directly released for QA purposes and they assume input as a formal text where the mention of entities is typically written in capital letters.}
Nevertheless, in case of QA systems, we cannot expect that a user writes questions with entities written in uppercase letters.
Previously these components relied on statistical approaches, however, with the support of knowledge graphs, we believe this drawback is easily fixable.
Similarly, the performance in case of questions having entities with long expressions can be enhanced by semantics and structure provided by KGs, such as the work presented in ~\cite{DBLP:conf/cikm/ReinandaMR16}.
The other limitation is recognising implicit entities. 19 out of 20 NED components received more than 50\% decrease in their performance for questions with one implicit entity. Furthermore, there are 210 questions which are not processed by any NED component; 85\% of them have implicit entities.
Since a given query is inherently short, it might not contain sufficient contextual information.
Recently, this problem was targeted in other research communities for social media~\cite{DBLP:conf/esws/PereraMAST16} and clinical documents~\cite{DBLP:conf/starsem/PereraMSTAHM15}, but little work related to QA was done.
Thus, we believe that the solutions employed in other research areas can inspire QA researchers. In addition, taking into account personalised information can support a better, context-aware QA system.
\end{inparaenum}
~\\
{\bf Suggestions to improve RL Components.} \begin{inparaenum}[\itshape i\upshape)] 
The two major pitfalls of RL components is the existence of implicit and hidden relations.
This challenge in RL task is more severe compared to the NED task because of the more profound complexity of relations (they can be expressed as a verb, adjective, adverb and noun phrase or even hidden in prepositions).
Lack of sufficient context also adds further complexity to relations. Since compared to RL components, existing techniques for NED are more stable, one possible improvement is to exploit entities extracted from the question to support relation linking (the existing RL components do not consider entities).
\end{inparaenum}
~\\
{\bf Suggestions to improve CL Components.} \begin{inparaenum}[\itshape i\upshape)] 
Similar to the two previous tasks, implicit mention of classes is a challenge.
We believe that approaches considering the hierarchy of the background ontology or taxonomies from external resources such as WordNet~\cite{DBLP:journals/cacm/Miller95} can improve CL components further.
\end{inparaenum}
~\\
{\bf Insight on QB Components.} \begin{inparaenum}[\itshape i\upshape)] 
As we mentioned previously, complex questions challenge QB components.
Typically, QB components rely on two methods, the first and popular method is template-based approaches (either using predefined templates or inferring templates from dependency parsers). The truth is that the more complex questions lead to more number of possible templates, therefore, obviously this influences performance negatively.
The second approach is traversal on KG which is commonly combined with light inference approaches. This method is not effective once the size of the underlying KG is large because, in case of long questions, it requires traversing a larger part of the KG. We recommend the research community to investigate new ways for implementing more efficient QB components.
\end{inparaenum}
~\\
{\bf Need of Frameworks for Micro Evaluation.} \begin{inparaenum}[\itshape i\upshape)] 
So far various benchmarking frameworks have been developed to provide analysis of strengths and weaknesses of existing open QA systems and associated components~\cite{GERBILQA,DBLP:conf/www/UsbeckRNBBBCCCE15}. However, they do not provide micro benchmarking of the components and systems and do not investigate the effect of question features in the related experiments. 
We contributed in providing a micro-level evaluation for performing such an analysis\textcolor{black}{, that is, we reused two renowned QA benchmarks for evaluating QA systems over the DBpedia KG and adapted them accordingly in order to evaluate the participating QA tasks separately}.
One of the limitations for such an analysis is attributed to the dataset.
The existing datasets do not have a fair distribution of all types of questions and question features.
\end{inparaenum}
~\\
{\bf Necessity to Address Vocabulary Mismatch Problem.} \begin{inparaenum}[\itshape i\upshape)] 
The implicit entities as well as implicit relations and classes have a negative impact on the performance of QA components.
Most issues are usually caused by vocabulary mismatch when the mention of entity/relation/class differs from its association in the background KG and schema.
This challenge is very important for schema-aware QA systems rather than schema-unaware search systems such as information retrieval approaches because the precise interpretation of the input query as well as the accurate spotting of the answer is more demanding~\cite{DBLP:conf/aaai/ShekarpourMAS17}.
Query expansion and rewriting are typical solutions for addressing the vocabulary mismatch problem. Thus, we recommend the QA research community to consider the development of components tackling vocabulary mismatch when implementing QA systems.
\end{inparaenum}
\section{Conclusions} \label{sec:conclusions}
In this paper, we performed an in-depth analysis of the performance of several QA components over DBpedia KG. The purpose of this analysis was to investigate three main research questions related to (i) the performance of QA components depending on different question features, (ii) the similarity in the behaviour of QA components based on different parameters and (iii) the impact of combining QA components on the overall performance respectively. \textcolor{black}{Rather than evaluating and discussing each QA component individually, we employ an overall evaluation approach which provides an overview and further insights in the successes and failures of existing QA components.} The necessity of this extended study emerged from the observations we have made on more than 60 QA systems and several other independent QA components which have been published until now. In fact, there is a lack in micro benchmarking of QA components addressing different QA-related tasks, with respect to input questions with different characteristics, which impedes the development of collaborative efforts to improve the state of the art.
With the current study we contribute to the research community with insightful results over 29 QA components based on 59 question features using a benchmark consisting of more than 3,000 questions from LC-QuAD, an analysis of pitfalls of existing QA components and their causes and, finally, a list of challenges and research directions in question answering. 

Although we focus on DBpedia, we believe that the results may generalise beyond this to other large cross-domain knowledge graphs, such as YAGO \cite{DBLP:conf/www/SuchanekKW07} and Wikidata \cite{DBLP:conf/www/Vrandecic12}. As most of the QA components are not tailored for DBpedia specifically\footnote{18 components evaluated in this study were not exclusively released for DBpedia, but were for both DBpedia and Wikipedia in general; some of them additionally provide dismabiguated URIs for Yago and Freebase.}, the QA pipelines for those graphs are similar, and the graphs share similar structures and overlap in content. It is important to extend this study with more QA components -- also components that are part of monolithic QA systems -- and with further benchmarks -- both domain-specific and domain-independent. Furthermore, our proposed analysis can be used to gain insights with regard to these or similar components beyond QA research.

One of the main lessons learned from this analysis is that none of the QA components per QA task is perfect but their performance varies based on questions with different features. \textcolor{black}{We conclude that representing performance of a QA component on average over all questions of a benchmarking dataset is not representative enough for analysing the components' strengths and does not shed light on their concrete weaknesses}. In particular, we found out that one of the main challenges for most components is the vocabulary mismatch problem (in NED, RL and CL task) and that in many cases, the appearance of uppercase letters in the entities (for NED) and the expected answer type (for QB) influence the performance of the corresponding components negatively.
Our ambition is that the results of this evaluation will encourage the QA research community to overcome the current drawbacks of the state of the art that prevent many of these approaches to be employed in real world applications.

\balance


\bibliographystyle{ACM-Reference-Format}
\bibliography{references}


\begin{thebibliography}{00}


\ifx \showCODEN    \undefined \def \showCODEN     #1{\unskip}     \fi
\ifx \showDOI      \undefined \def \showDOI       #1{#1}\fi
\ifx \showISBNx    \undefined \def \showISBNx     #1{\unskip}     \fi
\ifx \showISBNxiii \undefined \def \showISBNxiii  #1{\unskip}     \fi
\ifx \showISSN     \undefined \def \showISSN      #1{\unskip}     \fi
\ifx \showLCCN     \undefined \def \showLCCN      #1{\unskip}     \fi
\ifx \shownote     \undefined \def \shownote      #1{#1}          \fi
\ifx \showarticletitle \undefined \def \showarticletitle #1{#1}   \fi
\ifx \showURL      \undefined \def \showURL       {\relax}        \fi
\providecommand\bibfield[2]{#2}
\providecommand\bibinfo[2]{#2}
\providecommand\natexlab[1]{#1}
\providecommand\showeprint[2][]{arXiv:#2}

\bibitem[\protect\citeauthoryear{Auer, Bizer, Kobilarov, Lehmann, Cyganiak, and
  Ives}{Auer et~al\mbox{.}}{2007}]%
        {DBLP:conf/semweb/AuerBKLCI07}
\bibfield{author}{\bibinfo{person}{S{\"{o}}ren Auer},
  \bibinfo{person}{Christian Bizer}, \bibinfo{person}{Georgi Kobilarov},
  \bibinfo{person}{Jens Lehmann}, \bibinfo{person}{Richard Cyganiak}, {and}
  \bibinfo{person}{Zachary~G. Ives}.} \bibinfo{year}{2007}\natexlab{}.
\newblock \showarticletitle{{DBpedia: A Nucleus for a Web of Open Data}}. In
  \bibinfo{booktitle}{{\em {The Semantic Web, 6th International Semantic Web
  Conference, 2nd Asian Semantic Web Conference, {ISWC} 2007 + {ASWC} 2007,
  Busan, Korea, November 11-15, 2007.}}} \bibinfo{publisher}{Springer},
  \bibinfo{pages}{722--735}.
\newblock
\showDOI{%
\url{https://doi.org/10.1007/978-3-540-76298-0_52}}


\bibitem[\protect\citeauthoryear{Berant, Chou, Frostig, and Liang}{Berant
  et~al\mbox{.}}{2013}]%
        {DBLP:conf/emnlp/BerantCFL13}
\bibfield{author}{\bibinfo{person}{Jonathan Berant}, \bibinfo{person}{Andrew
  Chou}, \bibinfo{person}{Roy Frostig}, {and} \bibinfo{person}{Percy Liang}.}
  \bibinfo{year}{2013}\natexlab{}.
\newblock \showarticletitle{{Semantic Parsing on Freebase from Question-Answer
  Pairs}}. In \bibinfo{booktitle}{{\em {Proceedings of the 2013 Conference on
  Empirical Methods in Natural Language Processing, {EMNLP} 2013, 18-21 October
  2013, Grand Hyatt Seattle, Seattle, Washington, USA, {A} meeting of SIGDAT, a
  Special Interest Group of the {ACL}}}}. \bibinfo{publisher}{{ACL}},
  \bibinfo{pages}{1533--1544}.
\newblock
\showURL{%
\url{http://aclweb.org/anthology/D/D13/D13-1160.pdf}}


\bibitem[\protect\citeauthoryear{Bollacker, Evans, Paritosh, Sturge, and
  Taylor}{Bollacker et~al\mbox{.}}{2008}]%
        {DBLP:conf/aaai/BollackerCT07}
\bibfield{author}{\bibinfo{person}{Kurt~D. Bollacker}, \bibinfo{person}{Colin
  Evans}, \bibinfo{person}{Praveen Paritosh}, \bibinfo{person}{Tim Sturge},
  {and} \bibinfo{person}{Jamie Taylor}.} \bibinfo{year}{2008}\natexlab{}.
\newblock \showarticletitle{Freebase: a collaboratively created graph database
  for structuring human knowledge}. In \bibinfo{booktitle}{{\em {Proceedings of
  the {ACM} {SIGMOD} International Conference on Management of Data, {SIGMOD}
  2008, Vancouver, BC, Canada, June 10-12, 2008}}}. \bibinfo{publisher}{{ACM}},
  \bibinfo{pages}{1247--1250}.
\newblock
\showDOI{%
\url{https://doi.org/10.1145/1376616.1376746}}


\bibitem[\protect\citeauthoryear{Bordes, Usunier, Chopra, and Weston}{Bordes
  et~al\mbox{.}}{2015}]%
        {DBLP:journals/corr/BordesUCW15}
\bibfield{author}{\bibinfo{person}{Antoine Bordes}, \bibinfo{person}{Nicolas
  Usunier}, \bibinfo{person}{Sumit Chopra}, {and} \bibinfo{person}{Jason
  Weston}.} \bibinfo{year}{2015}\natexlab{}.
\newblock \showarticletitle{{Large-scale Simple Question Answering with Memory
  Networks}}.
\newblock \bibinfo{journal}{{\em CoRR\/}}  \bibinfo{volume}{abs/1506.02075}
  (\bibinfo{year}{2015}).
\newblock
\showeprint[arxiv]{1506.02075}
\showURL{%
\url{http://arxiv.org/abs/1506.02075}}


\bibitem[\protect\citeauthoryear{Cimiano, Lopez, Unger, Cabrio, Ngomo, and
  Walter}{Cimiano et~al\mbox{.}}{2013}]%
        {cimiano2013multilingual}
\bibfield{author}{\bibinfo{person}{Philipp Cimiano}, \bibinfo{person}{Vanessa
  Lopez}, \bibinfo{person}{Christina Unger}, \bibinfo{person}{Elena Cabrio},
  \bibinfo{person}{Axel-Cyrille~Ngonga Ngomo}, {and} \bibinfo{person}{Sebastian
  Walter}.} \bibinfo{year}{2013}\natexlab{}.
\newblock \showarticletitle{{Multilingual question answering over linked data
  (qald-3): Lab overview}}. In \bibinfo{booktitle}{{\em {International
  Conference of the Cross-Language Evaluation Forum for European Languages}}}.
  Springer, \bibinfo{pages}{321--332}.
\newblock


\bibitem[\protect\citeauthoryear{Cui, Xiao, Wang, Song, Hwang, and Wang}{Cui
  et~al\mbox{.}}{2017}]%
        {DBLP:journals/pvldb/CuiXWSHW17}
\bibfield{author}{\bibinfo{person}{Wanyun Cui}, \bibinfo{person}{Yanghua Xiao},
  \bibinfo{person}{Haixun Wang}, \bibinfo{person}{Yangqiu Song},
  \bibinfo{person}{Seung{-}won Hwang}, {and} \bibinfo{person}{Wei Wang}.}
  \bibinfo{year}{2017}\natexlab{}.
\newblock \showarticletitle{{KBQA: Learning Question Answering over QA Corpora
  and Knowledge Bases}}.
\newblock \bibinfo{journal}{{\em {PVLDB}\/}} \bibinfo{volume}{10},
  \bibinfo{number}{5} (\bibinfo{year}{2017}), \bibinfo{pages}{565--576}.
\newblock
\showURL{%
\url{http://www.vldb.org/pvldb/vol10/p565-cui.pdf}}


\bibitem[\protect\citeauthoryear{Dai, Li, and Xu}{Dai et~al\mbox{.}}{2016}]%
        {DBLP:conf/acl/DaiLX16}
\bibfield{author}{\bibinfo{person}{Zihang Dai}, \bibinfo{person}{Lei Li}, {and}
  \bibinfo{person}{Wei Xu}.} \bibinfo{year}{2016}\natexlab{}.
\newblock \showarticletitle{{CFO: Conditional Focused Neural Question Answering
  with Large-scale Knowledge Bases}}. In \bibinfo{booktitle}{{\em {Proceedings
  of the 54th Annual Meeting of the Association for Computational Linguistics,
  {ACL} 2016, August 7-12, 2016, Berlin, Germany, Volume 1: Long Papers}}}.
  \bibinfo{publisher}{The Association for Computer Linguistics}.
\newblock
\showURL{%
\url{http://aclweb.org/anthology/P/P16/P16-1076.pdf}}


\bibitem[\protect\citeauthoryear{Derczynski, Maynard, Rizzo, van Erp, Gorrell,
  Troncy, Petrak, and Bontcheva}{Derczynski et~al\mbox{.}}{2015}]%
        {DBLP:journals/ipm/DerczynskiM0EGTPB15}
\bibfield{author}{\bibinfo{person}{Leon Derczynski}, \bibinfo{person}{Diana
  Maynard}, \bibinfo{person}{Giuseppe Rizzo}, \bibinfo{person}{Marieke van
  Erp}, \bibinfo{person}{Genevieve Gorrell}, \bibinfo{person}{Rapha{\"{e}}l
  Troncy}, \bibinfo{person}{Johann Petrak}, {and} \bibinfo{person}{Kalina
  Bontcheva}.} \bibinfo{year}{2015}\natexlab{}.
\newblock \showarticletitle{Analysis of named entity recognition and linking
  for tweets}.
\newblock \bibinfo{journal}{{\em Inf. Process. Manage.\/}}
  \bibinfo{volume}{51}, \bibinfo{number}{2} (\bibinfo{year}{2015}),
  \bibinfo{pages}{32--49}.
\newblock


\bibitem[\protect\citeauthoryear{Diefenbach, Both, Singh, and Maret}{Diefenbach
  et~al\mbox{.}}{2018a}]%
        {diefenbach2018towards}
\bibfield{author}{\bibinfo{person}{Dennis Diefenbach}, \bibinfo{person}{Andreas
  Both}, \bibinfo{person}{Kamal Singh}, {and} \bibinfo{person}{Pierre Maret}.}
  \bibinfo{year}{2018}\natexlab{a}.
\newblock \showarticletitle{{Towards a Question Answering System over the
  Semantic Web}}.
\newblock \bibinfo{journal}{{\em arXiv preprint arXiv:1803.00832\/}}
  (\bibinfo{year}{2018}).
\newblock


\bibitem[\protect\citeauthoryear{Diefenbach, L{\'{o}}pez, Singh, and
  Maret}{Diefenbach et~al\mbox{.}}{2018b}]%
        {diefenbach2017core}
\bibfield{author}{\bibinfo{person}{Dennis Diefenbach}, \bibinfo{person}{Vanessa
  L{\'{o}}pez}, \bibinfo{person}{Kamal~Deep Singh}, {and}
  \bibinfo{person}{Pierre Maret}.} \bibinfo{year}{2018}\natexlab{b}.
\newblock \showarticletitle{Core techniques of question answering systems over
  knowledge bases: a survey}.
\newblock \bibinfo{journal}{{\em Knowledge and Information Systems\/}}
  \bibinfo{volume}{55}, \bibinfo{number}{3} (\bibinfo{year}{2018}),
  \bibinfo{pages}{529--569}.
\newblock
\showDOI{%
\url{https://doi.org/10.1007/s10115-017-1100-y}}


\bibitem[\protect\citeauthoryear{Ferragina and Scaiella}{Ferragina and
  Scaiella}{2010}]%
        {DBLP:conf/cikm/FerraginaS10}
\bibfield{author}{\bibinfo{person}{Paolo Ferragina} {and} \bibinfo{person}{Ugo
  Scaiella}.} \bibinfo{year}{2010}\natexlab{}.
\newblock \showarticletitle{{TAGME:} on-the-fly annotation of short text
  fragments (by wikipedia entities)}. In \bibinfo{booktitle}{{\em {Proceedings
  of the 19th ACM Conference on Information and Knowledge Management, CIKM
  2010, Toronto, Ontario, Canada, October 26-30, 2010}}}.
  \bibinfo{publisher}{{ACM}}, \bibinfo{pages}{1625--1628}.
\newblock
\showDOI{%
\url{https://doi.org/10.1145/1871437.1871689}}


\bibitem[\protect\citeauthoryear{Ferr{\'{a}}ndez, Spurk, Kouylekov, Dornescu,
  Ferr{\'{a}}ndez, Negri, Izquierdo, Tom{\'{a}}s, Orasan, Neumann, Magnini, and
  Gonz{\'{a}}lez}{Ferr{\'{a}}ndez et~al\mbox{.}}{2011}]%
        {qallme}
\bibfield{author}{\bibinfo{person}{{\'{O}}scar Ferr{\'{a}}ndez},
  \bibinfo{person}{Christian Spurk}, \bibinfo{person}{Milen Kouylekov},
  \bibinfo{person}{Iustin Dornescu}, \bibinfo{person}{Sergio Ferr{\'{a}}ndez},
  \bibinfo{person}{Matteo Negri}, \bibinfo{person}{Rub{\'{e}}n Izquierdo},
  \bibinfo{person}{David Tom{\'{a}}s}, \bibinfo{person}{Constantin Orasan},
  \bibinfo{person}{Guenter Neumann}, \bibinfo{person}{Bernardo Magnini}, {and}
  \bibinfo{person}{Jos{\'{e}} Luis~Vicedo Gonz{\'{a}}lez}.}
  \bibinfo{year}{2011}\natexlab{}.
\newblock \showarticletitle{{The QALL-ME Framework: A specifiable-domain
  multilingual Question Answering architecture}}.
\newblock \bibinfo{journal}{{\em J. Web Sem.\/}} \bibinfo{volume}{9},
  \bibinfo{number}{2} (\bibinfo{year}{2011}), \bibinfo{pages}{137--145}.
\newblock


\bibitem[\protect\citeauthoryear{Finkel, Grenager, and Manning}{Finkel
  et~al\mbox{.}}{2005}]%
        {DBLP:conf/acl/FinkelGM05}
\bibfield{author}{\bibinfo{person}{Jenny~Rose Finkel}, \bibinfo{person}{Trond
  Grenager}, {and} \bibinfo{person}{Christopher~D. Manning}.}
  \bibinfo{year}{2005}\natexlab{}.
\newblock \showarticletitle{{Incorporating Non-local Information into
  Information Extraction Systems by Gibbs Sampling}}. In
  \bibinfo{booktitle}{{\em {ACL 2005, 43rd Annual Meeting of the Association
  for Computational Linguistics, Proceedings of the Conference, 25-30 June
  2005, University of Michigan, USA}}}. \bibinfo{publisher}{The Association for
  Computer Linguistics}, \bibinfo{pages}{363--370}.
\newblock


\bibitem[\protect\citeauthoryear{Hoffart, Yosef, Bordino, F{\"{u}}rstenau,
  Pinkal, Spaniol, Taneva, Thater, and Weikum}{Hoffart et~al\mbox{.}}{2011}]%
        {DBLP:conf/emnlp/HoffartYBFPSTTW11}
\bibfield{author}{\bibinfo{person}{Johannes Hoffart},
  \bibinfo{person}{Mohamed~Amir Yosef}, \bibinfo{person}{Ilaria Bordino},
  \bibinfo{person}{Hagen F{\"{u}}rstenau}, \bibinfo{person}{Manfred Pinkal},
  \bibinfo{person}{Marc Spaniol}, \bibinfo{person}{Bilyana Taneva},
  \bibinfo{person}{Stefan Thater}, {and} \bibinfo{person}{Gerhard Weikum}.}
  \bibinfo{year}{2011}\natexlab{}.
\newblock \showarticletitle{{Robust Disambiguation of Named Entities in Text}}.
  In \bibinfo{booktitle}{{\em {Proceedings of the 2011 Conference on Empirical
  Methods in Natural Language Processing, EMNLP 2011, 27-31 July 2011, John
  McIntyre Conference Centre, Edinburgh, UK, A meeting of SIGDAT, a Special
  Interest Group of the ACL}}}. \bibinfo{publisher}{{ACL}},
  \bibinfo{pages}{782--792}.
\newblock
\showURL{%
\url{http://www.aclweb.org/anthology/D11-1072}}


\bibitem[\protect\citeauthoryear{Huang, Thint, and {\c{C}}elikyilmaz}{Huang
  et~al\mbox{.}}{2009}]%
        {DBLP:conf/emnlp/HuangTC09}
\bibfield{author}{\bibinfo{person}{Zhiheng Huang}, \bibinfo{person}{Marcus
  Thint}, {and} \bibinfo{person}{Asli {\c{C}}elikyilmaz}.}
  \bibinfo{year}{2009}\natexlab{}.
\newblock \showarticletitle{{Investigation of Question Classifier in Question
  Answering}}. In \bibinfo{booktitle}{{\em {Proceedings of the 2009 Conference
  on Empirical Methods in Natural Language Processing, {EMNLP} 2009, 6-7 August
  2009, Singapore, {A} meeting of SIGDAT, a Special Interest Group of the
  {ACL}}}}. \bibinfo{publisher}{{ACL}}, \bibinfo{pages}{543--550}.
\newblock


\bibitem[\protect\citeauthoryear{Kim, Choi, Kim, Kim, and Choi}{Kim
  et~al\mbox{.}}{2016}]%
        {kim2016open}
\bibfield{author}{\bibinfo{person}{Jiseong Kim}, \bibinfo{person}{GyuHyeon
  Choi}, \bibinfo{person}{Jung-Uk Kim}, \bibinfo{person}{Eun-Kyung Kim}, {and}
  \bibinfo{person}{Key-Sun Choi}.} \bibinfo{year}{2016}\natexlab{}.
\newblock \showarticletitle{{The Open Framework for Developing Knowledge Base
  And Question Answering System}}. In \bibinfo{booktitle}{{\em {Proceedings of
  COLING 2016, the 26th International Conference on Computational
  Linguistics}}}. \bibinfo{publisher}{{ACL}}, \bibinfo{pages}{161--165}.
\newblock


\bibitem[\protect\citeauthoryear{Li and Jagadish}{Li and Jagadish}{2014}]%
        {DBLP:journals/pvldb/LiJ14}
\bibfield{author}{\bibinfo{person}{Fei Li} {and} \bibinfo{person}{H.~V.
  Jagadish}.} \bibinfo{year}{2014}\natexlab{}.
\newblock \showarticletitle{{Constructing an Interactive Natural Language
  Interface for Relational Databases}}.
\newblock \bibinfo{journal}{{\em {PVLDB}\/}} \bibinfo{volume}{8},
  \bibinfo{number}{1} (\bibinfo{year}{2014}), \bibinfo{pages}{73--84}.
\newblock
\showURL{%
\url{http://www.vldb.org/pvldb/vol8/p73-li.pdf}}


\bibitem[\protect\citeauthoryear{Marx, Usbeck, Ngomo, H{\"{o}}ffner, Lehmann,
  and Auer}{Marx et~al\mbox{.}}{2014}]%
        {openqa}
\bibfield{author}{\bibinfo{person}{Edgard Marx}, \bibinfo{person}{Ricardo
  Usbeck}, \bibinfo{person}{Axel{-}Cyrille~Ngonga Ngomo},
  \bibinfo{person}{Konrad H{\"{o}}ffner}, \bibinfo{person}{Jens Lehmann}, {and}
  \bibinfo{person}{S{\"{o}}ren Auer}.} \bibinfo{year}{2014}\natexlab{}.
\newblock \showarticletitle{Towards an open question answering architecture}.
  In \bibinfo{booktitle}{{\em {Proceedings of the 10th International Conference
  on Semantic Systems, {SEMANTICS} 2014, Leipzig, Germany, September 4-5,
  2014}}}. \bibinfo{publisher}{{ACM}}, \bibinfo{pages}{57--60}.
\newblock


\bibitem[\protect\citeauthoryear{Mazzeo and Zaniolo}{Mazzeo and
  Zaniolo}{2016}]%
        {DBLP:conf/edbt/MazzeoZ16}
\bibfield{author}{\bibinfo{person}{Giuseppe~M. Mazzeo} {and}
  \bibinfo{person}{Carlo Zaniolo}.} \bibinfo{year}{2016}\natexlab{}.
\newblock \showarticletitle{{Answering Controlled Natural Language Questions on
  RDF Knowledge Bases}}. In \bibinfo{booktitle}{{\em {Proceedings of the 19th
  International Conference on Extending Database Technology, {EDBT} 2016,
  Bordeaux, France, March 15-16, 2016, Bordeaux, France, March 15-16, 2016.}}}
  \bibinfo{publisher}{OpenProceedings.org}, \bibinfo{pages}{608--611}.
\newblock
\showDOI{%
\url{https://doi.org/10.5441/002/edbt.2016.60}}


\bibitem[\protect\citeauthoryear{Mendes, Jakob, Garc{\'{\i}}a{-}Silva, and
  Bizer}{Mendes et~al\mbox{.}}{2011}]%
        {MendesJGB11}
\bibfield{author}{\bibinfo{person}{Pablo~N. Mendes}, \bibinfo{person}{Max
  Jakob}, \bibinfo{person}{Andr{\'{e}}s Garc{\'{\i}}a{-}Silva}, {and}
  \bibinfo{person}{Christian Bizer}.} \bibinfo{year}{2011}\natexlab{}.
\newblock \showarticletitle{{DBpedia spotlight: shedding light on the web of
  documents}}. In \bibinfo{booktitle}{{\em {Proceedings the 7th International
  Conference on Semantic Systems, {I-SEMANTICS} 2011, Graz, Austria, September
  7-9, 2011}}}. \bibinfo{publisher}{{ACM}}, \bibinfo{pages}{1--8}.
\newblock
\showDOI{%
\url{https://doi.org/10.1145/2063518.2063519}}


\bibitem[\protect\citeauthoryear{Miller}{Miller}{1995}]%
        {DBLP:journals/cacm/Miller95}
\bibfield{author}{\bibinfo{person}{George~A. Miller}.}
  \bibinfo{year}{1995}\natexlab{}.
\newblock \showarticletitle{{WordNet: A Lexical Database for English}}.
\newblock \bibinfo{journal}{{\em Commun. {ACM}\/}} \bibinfo{volume}{38},
  \bibinfo{number}{11} (\bibinfo{year}{1995}), \bibinfo{pages}{39--41}.
\newblock
\showDOI{%
\url{https://doi.org/10.1145/219717.219748}}


\bibitem[\protect\citeauthoryear{Moro, Raganato, and Navigli}{Moro
  et~al\mbox{.}}{2014}]%
        {DBLP:journals/tacl/0001RN14}
\bibfield{author}{\bibinfo{person}{Andrea Moro}, \bibinfo{person}{Alessandro
  Raganato}, {and} \bibinfo{person}{Roberto Navigli}.}
  \bibinfo{year}{2014}\natexlab{}.
\newblock \showarticletitle{{Entity Linking meets Word Sense Disambiguation: a
  Unified Approach}}.
\newblock \bibinfo{journal}{{\em {TACL}\/}}  \bibinfo{volume}{2}
  (\bibinfo{year}{2014}), \bibinfo{pages}{231--244}.
\newblock
\showURL{%
\url{https://tacl2013.cs.columbia.edu/ojs/index.php/tacl/article/view/291}}


\bibitem[\protect\citeauthoryear{Mulang, Singh, and Orlandi}{Mulang
  et~al\mbox{.}}{2017}]%
        {DBLP:conf/i-semantics/MulangSO17}
\bibfield{author}{\bibinfo{person}{Isaiah~Onando Mulang},
  \bibinfo{person}{Kuldeep Singh}, {and} \bibinfo{person}{Fabrizio Orlandi}.}
  \bibinfo{year}{2017}\natexlab{}.
\newblock \showarticletitle{{Matching Natural Language Relations to Knowledge
  Graph Properties for Question Answering}}. In \bibinfo{booktitle}{{\em
  {Proceedings of the 13th International Conference on Semantic Systems,
  {SEMANTICS} 2017, Amsterdam, The Netherlands, September 11-14, 2017}}}.
  \bibinfo{publisher}{{ACM}}, \bibinfo{pages}{89--96}.
\newblock


\bibitem[\protect\citeauthoryear{Perera, Mendes, Alex, Sheth, and
  Thirunarayan}{Perera et~al\mbox{.}}{2016}]%
        {DBLP:conf/esws/PereraMAST16}
\bibfield{author}{\bibinfo{person}{Sujan Perera}, \bibinfo{person}{Pablo~N.
  Mendes}, \bibinfo{person}{Adarsh Alex}, \bibinfo{person}{Amit~P. Sheth},
  {and} \bibinfo{person}{Krishnaprasad Thirunarayan}.}
  \bibinfo{year}{2016}\natexlab{}.
\newblock \showarticletitle{{Implicit Entity Linking in Tweets}}. In
  \bibinfo{booktitle}{{\em {The Semantic Web. Latest Advances and New Domains -
  13th International Conference, {ESWC} 2016, Heraklion, Crete, Greece, May 29
  - June 2, 2016, Proceedings}}}. \bibinfo{publisher}{Springer},
  \bibinfo{pages}{118--132}.
\newblock
\showDOI{%
\url{https://doi.org/10.1007/978-3-319-34129-3_8}}


\bibitem[\protect\citeauthoryear{Perera, Mendes, Sheth, Thirunarayan, Alex,
  Heid, and Mott}{Perera et~al\mbox{.}}{2015}]%
        {DBLP:conf/starsem/PereraMSTAHM15}
\bibfield{author}{\bibinfo{person}{Sujan Perera}, \bibinfo{person}{Pablo~N.
  Mendes}, \bibinfo{person}{Amit~P. Sheth}, \bibinfo{person}{Krishnaprasad
  Thirunarayan}, \bibinfo{person}{Adarsh Alex}, \bibinfo{person}{Christopher
  Heid}, {and} \bibinfo{person}{Greg Mott}.} \bibinfo{year}{2015}\natexlab{}.
\newblock \showarticletitle{{Implicit Entity Recognition in Clinical
  Documents}}. In \bibinfo{booktitle}{{\em {Proceedings of the Fourth Joint
  Conference on Lexical and Computational Semantics, *SEM 2015, June 4-5, 2015,
  Denver, Colorado, USA.}}} \bibinfo{publisher}{The *SEM 2015 Organizing
  Committee}, \bibinfo{pages}{228--238}.
\newblock
\showURL{%
\url{http://aclweb.org/anthology/S/S15/S15-1028.pdf}}


\bibitem[\protect\citeauthoryear{Reinanda, Meij, and de~Rijke}{Reinanda
  et~al\mbox{.}}{2016}]%
        {DBLP:conf/cikm/ReinandaMR16}
\bibfield{author}{\bibinfo{person}{Ridho Reinanda}, \bibinfo{person}{Edgar
  Meij}, {and} \bibinfo{person}{Maarten de Rijke}.}
  \bibinfo{year}{2016}\natexlab{}.
\newblock \showarticletitle{{Document Filtering for Long-tail Entities}}. In
  \bibinfo{booktitle}{{\em {Proceedings of the 25th ACM International
  Conference on Information and Knowledge Management, CIKM 2016, Indianapolis,
  IN, USA, October 24-28, 2016}}}. \bibinfo{publisher}{{ACM}},
  \bibinfo{pages}{771--780}.
\newblock
\showDOI{%
\url{https://doi.org/10.1145/2983323.2983728}}


\bibitem[\protect\citeauthoryear{Rousseeuw}{Rousseeuw}{1987}]%
        {Rousseeuw:1987:SGA:38768.38772}
\bibfield{author}{\bibinfo{person}{Peter Rousseeuw}.}
  \bibinfo{year}{1987}\natexlab{}.
\newblock \showarticletitle{{Silhouettes: A Graphical Aid to the Interpretation
  and Validation of Cluster Analysis}}.
\newblock \bibinfo{journal}{{\em J. Comput. Appl. Math.\/}}
  \bibinfo{volume}{20}, \bibinfo{number}{1} (\bibinfo{date}{Nov.}
  \bibinfo{year}{1987}), \bibinfo{pages}{53--65}.
\newblock
\showISSN{0377-0427}
\showDOI{%
\url{https://doi.org/10.1016/0377-0427(87)90125-7}}


\bibitem[\protect\citeauthoryear{Saleem, Dastjerdi, Usbeck, and Ngomo}{Saleem
  et~al\mbox{.}}{2017}]%
        {saleemquestion}
\bibfield{author}{\bibinfo{person}{Muhammad Saleem},
  \bibinfo{person}{Samaneh~Nazari Dastjerdi}, \bibinfo{person}{Ricardo Usbeck},
  {and} \bibinfo{person}{Axel{-}Cyrille~Ngonga Ngomo}.}
  \bibinfo{year}{2017}\natexlab{}.
\newblock \showarticletitle{{Question Answering Over Linked Data: What is
  Difficult to Answer? What Affects the F scores?}}. In
  \bibinfo{booktitle}{{\em {Joint Proceedings of {BLINK2017:} 2nd International
  Workshop on Benchmarking Linked Data and NLIWoD3: Natural Language Interfaces
  for the Web of Data co-located with 16th International Semantic Web
  Conference {(ISWC} 2017), Vienna, Austria, October 21st - to - 22nd, 2017.}}}
  \bibinfo{publisher}{CEUR-WS.org}.
\newblock
\showURL{%
\url{http://ceur-ws.org/Vol-1932/paper-02.pdf}}


\bibitem[\protect\citeauthoryear{Shekarpour, Marx, Auer, and Sheth}{Shekarpour
  et~al\mbox{.}}{2017}]%
        {DBLP:conf/aaai/ShekarpourMAS17}
\bibfield{author}{\bibinfo{person}{Saeedeh Shekarpour}, \bibinfo{person}{Edgard
  Marx}, \bibinfo{person}{S{\"{o}}ren Auer}, {and} \bibinfo{person}{Amit~P.
  Sheth}.} \bibinfo{year}{2017}\natexlab{}.
\newblock \showarticletitle{{RQUERY: Rewriting Natural Language Queries on
  Knowledge Graphs to Alleviate the Vocabulary Mismatch Problem}}. In
  \bibinfo{booktitle}{{\em {Proceedings of the Thirty-First {AAAI} Conference
  on Artificial Intelligence, February 4-9, 2017, San Francisco, California,
  USA.}}} \bibinfo{publisher}{{AAAI} Press}, \bibinfo{pages}{3936--3943}.
\newblock
\showURL{%
\url{http://aaai.org/ocs/index.php/AAAI/AAAI17/paper/view/14638}}


\bibitem[\protect\citeauthoryear{Shekarpour, Marx, Ngomo, and Auer}{Shekarpour
  et~al\mbox{.}}{2015}]%
        {DBLP:journals/ws/ShekarpourMNA15}
\bibfield{author}{\bibinfo{person}{Saeedeh Shekarpour}, \bibinfo{person}{Edgard
  Marx}, \bibinfo{person}{Axel{-}Cyrille~Ngonga Ngomo}, {and}
  \bibinfo{person}{S{\"{o}}ren Auer}.} \bibinfo{year}{2015}\natexlab{}.
\newblock \showarticletitle{{SINA: Semantic interpretation of user queries for
  question answering on interlinked data}}.
\newblock \bibinfo{journal}{{\em J. Web Sem.\/}}  \bibinfo{volume}{30}
  (\bibinfo{year}{2015}), \bibinfo{pages}{39--51}.
\newblock
\showDOI{%
\url{https://doi.org/10.1016/j.websem.2014.06.002}}


\bibitem[\protect\citeauthoryear{Singh, Both, Radhakrishna, and
  Shekarpour}{Singh et~al\mbox{.}}{2018a}]%
        {DBLP:conf/esws/SinghBRS18}
\bibfield{author}{\bibinfo{person}{Kuldeep Singh}, \bibinfo{person}{Andreas
  Both}, \bibinfo{person}{Arun~Sethupat Radhakrishna}, {and}
  \bibinfo{person}{Saeedeh Shekarpour}.} \bibinfo{year}{2018}\natexlab{a}.
\newblock \showarticletitle{{Frankenstein: A Platform Enabling Reuse of
  Question Answering Components}}. In \bibinfo{booktitle}{{\em {The Semantic
  Web - 15th International Conference, {ESWC} 2018, Heraklion, Crete, Greece,
  June 3-7, 2018, Proceedings}}}. \bibinfo{publisher}{Springer},
  \bibinfo{pages}{624--638}.
\newblock
\showDOI{%
\url{https://doi.org/10.1007/978-3-319-93417-4_40}}


\bibitem[\protect\citeauthoryear{Singh, Mulang, Lytra, Jaradeh, Sakor, Vidal,
  Lange, and Auer}{Singh et~al\mbox{.}}{2017}]%
        {kcap}
\bibfield{author}{\bibinfo{person}{Kuldeep Singh},
  \bibinfo{person}{Isaiah~Onando Mulang}, \bibinfo{person}{Ioanna Lytra},
  \bibinfo{person}{Mohamad~Yaser Jaradeh}, \bibinfo{person}{Ahmad Sakor},
  \bibinfo{person}{Maria{-}Esther Vidal}, \bibinfo{person}{Christoph Lange},
  {and} \bibinfo{person}{S{\"{o}}ren Auer}.} \bibinfo{year}{2017}\natexlab{}.
\newblock \showarticletitle{{Capturing Knowledge in Semantically-typed
  Relational Patterns to Enhance Relation Linking}}. In
  \bibinfo{booktitle}{{\em {Proceedings of the Knowledge Capture Conference,
  {K-CAP} 2017, Austin, TX, USA, December 4-6, 2017}}}.
  \bibinfo{publisher}{{ACM}}, \bibinfo{pages}{31:1--31:8}.
\newblock


\bibitem[\protect\citeauthoryear{Singh, Radhakrishna, Both, Shekarpour, Lytra,
  Usbeck, Vyas, Khikmatullaev, Punjani, Lange, Vidal, Lehmann, and Auer}{Singh
  et~al\mbox{.}}{2018b}]%
        {DBLP:conf/www/SinghRBSLUVKP0V18}
\bibfield{author}{\bibinfo{person}{Kuldeep Singh},
  \bibinfo{person}{Arun~Sethupat Radhakrishna}, \bibinfo{person}{Andreas Both},
  \bibinfo{person}{Saeedeh Shekarpour}, \bibinfo{person}{Ioanna Lytra},
  \bibinfo{person}{Ricardo Usbeck}, \bibinfo{person}{Akhilesh Vyas},
  \bibinfo{person}{Akmal Khikmatullaev}, \bibinfo{person}{Dharmen Punjani},
  \bibinfo{person}{Christoph Lange}, \bibinfo{person}{Maria{-}Esther Vidal},
  \bibinfo{person}{Jens Lehmann}, {and} \bibinfo{person}{S{\"{o}}ren Auer}.}
  \bibinfo{year}{2018}\natexlab{b}.
\newblock \showarticletitle{{Why Reinvent the Wheel: Let's Build Question
  Answering Systems Together}}. In \bibinfo{booktitle}{{\em {Proceedings of the
  2018 World Wide Web Conference, {WWW} 2018, Lyon, France, April 23-27,
  2018}}}. \bibinfo{publisher}{{ACM}}, \bibinfo{pages}{1247--1256}.
\newblock
\showDOI{%
\url{https://doi.org/10.1145/3178876.3186023}}


\bibitem[\protect\citeauthoryear{Suchanek, Kasneci, and Weikum}{Suchanek
  et~al\mbox{.}}{2007}]%
        {DBLP:conf/www/SuchanekKW07}
\bibfield{author}{\bibinfo{person}{Fabian~M. Suchanek},
  \bibinfo{person}{Gjergji Kasneci}, {and} \bibinfo{person}{Gerhard Weikum}.}
  \bibinfo{year}{2007}\natexlab{}.
\newblock \showarticletitle{Yago: a core of semantic knowledge}. In
  \bibinfo{booktitle}{{\em {Proceedings of the 16th International Conference on
  World Wide Web, {WWW} 2007, Banff, Alberta, Canada, May 8-12, 2007}}}.
  \bibinfo{publisher}{{ACM}}, \bibinfo{pages}{697--706}.
\newblock
\showDOI{%
\url{https://doi.org/10.1145/1242572.1242667}}


\bibitem[\protect\citeauthoryear{Trivedi, Maheshwari, Dubey, and
  Lehmann}{Trivedi et~al\mbox{.}}{2017}]%
        {trivedi2017lc}
\bibfield{author}{\bibinfo{person}{Priyansh Trivedi}, \bibinfo{person}{Gaurav
  Maheshwari}, \bibinfo{person}{Mohnish Dubey}, {and} \bibinfo{person}{Jens
  Lehmann}.} \bibinfo{year}{2017}\natexlab{}.
\newblock \showarticletitle{{LC-QuAD: A Corpus for Complex Question Answering
  over Knowledge Graphs}}. In \bibinfo{booktitle}{{\em {The Semantic Web -
  {ISWC} 2017 - 16th International Semantic Web Conference, Vienna, Austria,
  October 21-25, 2017, Proceedings, Part II}}}. \bibinfo{publisher}{Springer},
  \bibinfo{pages}{210--218}.
\newblock
\showDOI{%
\url{https://doi.org/10.1007/978-3-319-68204-4_22}}


\bibitem[\protect\citeauthoryear{Unger, Forascu, L{\'{o}}pez, Ngomo, Cabrio,
  Cimiano, and Walter}{Unger et~al\mbox{.}}{2015}]%
        {DBLP:conf/clef/UngerFLNCCW15}
\bibfield{author}{\bibinfo{person}{Christina Unger}, \bibinfo{person}{Corina
  Forascu}, \bibinfo{person}{Vanessa L{\'{o}}pez},
  \bibinfo{person}{Axel{-}Cyrille~Ngonga Ngomo}, \bibinfo{person}{Elena
  Cabrio}, \bibinfo{person}{Philipp Cimiano}, {and} \bibinfo{person}{Sebastian
  Walter}.} \bibinfo{year}{2015}\natexlab{}.
\newblock \showarticletitle{{Question Answering over Linked Data (QALD-5)}}. In
  \bibinfo{booktitle}{{\em {Working Notes of {CLEF} 2015 - Conference and Labs
  of the Evaluation forum, Toulouse, France, September 8-11, 2015.}}}
  \bibinfo{publisher}{CEUR-WS.org}.
\newblock
\showURL{%
\url{http://ceur-ws.org/Vol-1391/173-CR.pdf}}


\bibitem[\protect\citeauthoryear{Usbeck, Ngomo, R{\"{o}}der, Gerber, Coelho,
  Auer, and Both}{Usbeck et~al\mbox{.}}{2014}]%
        {UsbeckNRGCAB14}
\bibfield{author}{\bibinfo{person}{Ricardo Usbeck},
  \bibinfo{person}{Axel{-}Cyrille~Ngonga Ngomo}, \bibinfo{person}{Michael
  R{\"{o}}der}, \bibinfo{person}{Daniel Gerber},
  \bibinfo{person}{Sandro~Athaide Coelho}, \bibinfo{person}{S{\"{o}}ren Auer},
  {and} \bibinfo{person}{Andreas Both}.} \bibinfo{year}{2014}\natexlab{}.
\newblock \showarticletitle{{AGDISTIS - Graph-Based Disambiguation of Named
  Entities Using Linked Data}}. In \bibinfo{booktitle}{{\em {The Semantic Web -
  {ISWC} 2014}}}. \bibinfo{publisher}{Springer}, \bibinfo{pages}{457--471}.
\newblock
\showDOI{%
\url{https://doi.org/10.1007/978-3-319-11964-9_29}}


\bibitem[\protect\citeauthoryear{Usbeck, R\"oder, Hoffmann, Conrads, Huthmann,
  Ngomo, Demmler, and Unger}{Usbeck et~al\mbox{.}}{2017}]%
        {GERBILQA}
\bibfield{author}{\bibinfo{person}{Ricardo Usbeck}, \bibinfo{person}{Michael
  R\"oder}, \bibinfo{person}{Michael Hoffmann}, \bibinfo{person}{Felix
  Conrads}, \bibinfo{person}{Jonathan Huthmann},
  \bibinfo{person}{Axel-Cyrille~Ngonga Ngomo}, \bibinfo{person}{Christian
  Demmler}, {and} \bibinfo{person}{Christina Unger}.}
  \bibinfo{year}{2017}\natexlab{}.
\newblock \showarticletitle{{Benchmarking Question Answering Systems}}.
\newblock \bibinfo{journal}{{\em {Semantic Web Journal}\/}}
  (\bibinfo{year}{2017}).
\newblock
\showURL{%
\url{http://www.semantic-web-journal.net/content/benchmarking-question-answering-systems}}


\bibitem[\protect\citeauthoryear{Usbeck, R{\"{o}}der, Ngomo, Baron, Both,
  Br{\"{u}}mmer, Ceccarelli, Cornolti, Cherix, Eickmann, Ferragina, Lemke,
  Moro, Navigli, Piccinno, Rizzo, Sack, Speck, Troncy, Waitelonis, and
  Wesemann}{Usbeck et~al\mbox{.}}{2015}]%
        {DBLP:conf/www/UsbeckRNBBBCCCE15}
\bibfield{author}{\bibinfo{person}{Ricardo Usbeck}, \bibinfo{person}{Michael
  R{\"{o}}der}, \bibinfo{person}{Axel{-}Cyrille~Ngonga Ngomo},
  \bibinfo{person}{Ciro Baron}, \bibinfo{person}{Andreas Both},
  \bibinfo{person}{Martin Br{\"{u}}mmer}, \bibinfo{person}{Diego Ceccarelli},
  \bibinfo{person}{Marco Cornolti}, \bibinfo{person}{Didier Cherix},
  \bibinfo{person}{Bernd Eickmann}, \bibinfo{person}{Paolo Ferragina},
  \bibinfo{person}{Christiane Lemke}, \bibinfo{person}{Andrea Moro},
  \bibinfo{person}{Roberto Navigli}, \bibinfo{person}{Francesco Piccinno},
  \bibinfo{person}{Giuseppe Rizzo}, \bibinfo{person}{Harald Sack},
  \bibinfo{person}{Ren{\'{e}} Speck}, \bibinfo{person}{Rapha{\"{e}}l Troncy},
  \bibinfo{person}{J{\"{o}}rg Waitelonis}, {and} \bibinfo{person}{Lars
  Wesemann}.} \bibinfo{year}{2015}\natexlab{}.
\newblock \showarticletitle{{GERBIL: General Entity Annotator Benchmarking
  Framework}}. In \bibinfo{booktitle}{{\em {Proceedings of the 24th
  International Conference on World Wide Web, {WWW} 2015, Florence, Italy, May
  18-22, 2015}}}. \bibinfo{publisher}{{ACM}}, \bibinfo{pages}{1133--1143}.
\newblock
\showDOI{%
\url{https://doi.org/10.1145/2736277.2741626}}


\bibitem[\protect\citeauthoryear{Voorhees and Harman}{Voorhees and
  Harman}{2000}]%
        {voorhees2000overview}
\bibfield{author}{\bibinfo{person}{Ellen~M Voorhees} {and}
  \bibinfo{person}{Donna Harman}.} \bibinfo{year}{2000}\natexlab{}.
\newblock \showarticletitle{{Overview of the sixth text retrieval conference
  (TREC-6)}}.
\newblock \bibinfo{journal}{{\em Information Processing \& Management\/}}
  \bibinfo{volume}{36}, \bibinfo{number}{1} (\bibinfo{year}{2000}),
  \bibinfo{pages}{3--35}.
\newblock


\bibitem[\protect\citeauthoryear{Vrandecic}{Vrandecic}{2012}]%
        {DBLP:conf/www/Vrandecic12}
\bibfield{author}{\bibinfo{person}{Denny Vrandecic}.}
  \bibinfo{year}{2012}\natexlab{}.
\newblock \showarticletitle{Wikidata: a new platform for collaborative data
  collection}. In \bibinfo{booktitle}{{\em {Proceedings of the 21st World Wide
  Web Conference, {WWW} 2012, Lyon, France, April 16-20, 2012 (Companion
  Volume)}}}. \bibinfo{publisher}{{ACM}}, \bibinfo{pages}{1063--1064}.
\newblock
\showDOI{%
\url{https://doi.org/10.1145/2187980.2188242}}


\bibitem[\protect\citeauthoryear{Yu, Yin, Hasan, dos Santos, Xiang, and
  Zhou}{Yu et~al\mbox{.}}{2017}]%
        {DBLP:conf/acl/YuYHSXZ17}
\bibfield{author}{\bibinfo{person}{Mo Yu}, \bibinfo{person}{Wenpeng Yin},
  \bibinfo{person}{Kazi~Saidul Hasan}, \bibinfo{person}{C{\'{\i}}cero~Nogueira
  dos Santos}, \bibinfo{person}{Bing Xiang}, {and} \bibinfo{person}{Bowen
  Zhou}.} \bibinfo{year}{2017}\natexlab{}.
\newblock \showarticletitle{{Improved Neural Relation Detection for Knowledge
  Base Question Answering}}. In \bibinfo{booktitle}{{\em {Proceedings of the
  55th Annual Meeting of the Association for Computational Linguistics, {ACL}
  2017, Vancouver, Canada, July 30 - August 4, Volume 1: Long Papers}}}.
  \bibinfo{publisher}{Association for Computational Linguistics},
  \bibinfo{pages}{571--581}.
\newblock
\showDOI{%
\url{https://doi.org/10.18653/v1/P17-1053}}


\bibitem[\protect\citeauthoryear{Zou, Huang, Wang, Yu, He, and Zhao}{Zou
  et~al\mbox{.}}{2014}]%
        {DBLP:conf/sigmod/ZouHWYHZ14}
\bibfield{author}{\bibinfo{person}{Lei Zou}, \bibinfo{person}{Ruizhe Huang},
  \bibinfo{person}{Haixun Wang}, \bibinfo{person}{Jeffrey~Xu Yu},
  \bibinfo{person}{Wenqiang He}, {and} \bibinfo{person}{Dongyan Zhao}.}
  \bibinfo{year}{2014}\natexlab{}.
\newblock \showarticletitle{Natural language question answering over {RDF:} a
  graph data driven approach}. In \bibinfo{booktitle}{{\em {International
  Conference on Management of Data, {SIGMOD} 2014, Snowbird, UT, USA, June
  22-27, 2014}}}. \bibinfo{publisher}{{ACM}}, \bibinfo{pages}{313--324}.
\newblock
\showDOI{%
\url{https://doi.org/10.1145/2588555.2610525}}


\end{thebibliography}

\end{document}